\documentclass[prd,
reprint,
superscriptaddress,
preprintnumbers,
nofootinbib,
 amsmath,amssymb,
 aps,
]{revtex4-2}

\usepackage[english]{babel}
\usepackage{xcolor}
\usepackage[
  linktocpage,
  colorlinks,
  citecolor=blue,
  linkcolor=blue,
  urlcolor=blue
]{hyperref}

\usepackage{enumerate}
\usepackage{float}
\usepackage{calc}
\usepackage{comment}
\usepackage{upgreek}
\usepackage{IEEEtrantools}
\usepackage{Commands}
\usepackage{soul}
\usepackage[normalem]{ulem}
\usepackage{physics}

\usepackage{caption}
\usepackage{subcaption}
\usepackage{graphicx}
\usepackage{dcolumn}
\usepackage{bm}

\DeclareRobustCommand{\pzcletter}[1]{%
  \mathord{%
    \mathchoice
      {\text{{\fontsize{11}{11}\selectfont\fontfamily{pzc}\fontshape{it}\selectfont #1}}}%
      {\text{{\fontsize{11}{11}\selectfont\fontfamily{pzc}\fontshape{it}\selectfont #1}}}%
      {\text{{\fontsize{9}{9}\selectfont\fontfamily{pzc}\fontshape{it}\selectfont #1}}}%
      {\text{{\fontsize{7}{7}\selectfont\fontfamily{pzc}\fontshape{it}\selectfont #1}}}%
  }%
}
\newcommand{\mscr}{\pzcletter{m}}

\newcommand{\orcid}[1]{\href{https://orcid.org/#1}{\includegraphics[width=8pt]{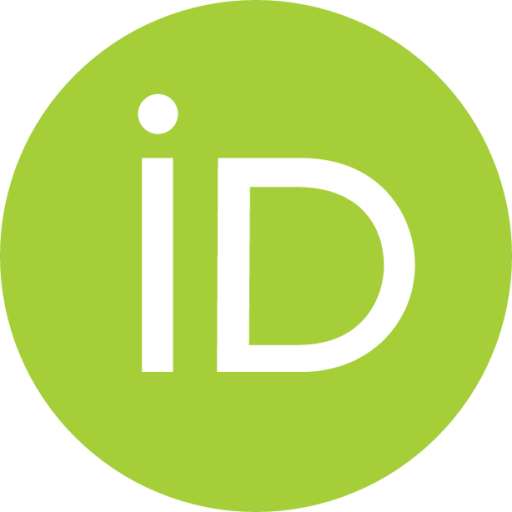}}}

\newcommand{\jz}[1]{\textcolor{orange}{[jz: #1]}}

\newcommand{\sg}[1]{\textcolor{purple}{[sg: #1]}}

\newcommand{\las}[1]{\textcolor{red}{[las: #1]}}

\newcommand{\DDD}[1]{\textcolor{brown}{[dd: #1]}}

\begin{document}

\preprint{CERN-TH-2026-086}

\title{Gravitational Memory from Hairy Binary Black Hole Mergers}
\author{Silvia Gasparotto~\orcid{0000-0001-7586-1786}}
\email{silvia.gasparotto@cern.ch}
\affiliation{CERN, Theoretical Physics Department, Esplanade des Particules 1, Geneva 1211, Switzerland}

\author{Jann Zosso~\orcid{0000-0002-2671-7531}}
\email{jann.zosso@nbi.ku.dk}
\affiliation{Center of Gravity, Niels Bohr Institute, Blegdamsvej 17, 2100 Copenhagen, Denmark}

\author{Llibert Aresté Saló~\orcid{0000-0002-3812-8523}}
\email{llibert.arestesalo@kuleuven.be}
\affiliation{Instituut voor Theoretische Fysica, KU Leuven. Celestijnenlaan 200D, B-3001 Leuven, Belgium}
\affiliation{Leuven Gravity Institute, KU Leuven. Celestijnenlaan 200D, B-3001 Leuven, Belgium}

\author{Daniela D. Doneva~\orcid{0000-0001-6519-000X}}
\email{daniela.doneva@uv.es}
\affiliation{Departamento de Astronom\'ia y Astrof\'isica, Universitat de Val\`encia,
Av. Vicent Andrés Estellés, 19, 46100, Burjassot (Val\`encia), Spain}
\affiliation{Theoretical Astrophysics, Eberhard Karls University of T\"ubingen, 72076 T\"ubingen, Germany}

\author{Stoytcho S. Yazadjiev~\orcid{0000-0002-1280-9013}  }
\email{yazad@phys.uni-sofia.bg}
\affiliation{Department of Theoretical Physics, Sofia University ``St. Kliment Ohridski", Sofia 1164, Bulgaria}
\affiliation{Institute of Mathematics and Informatics, Bulgarian Academy of Sciences, Acad. G. Bonchev St. 8, Sofia 1113, Bulgaria}

\date{\today}

\begin{abstract}
Gravitational-wave memory is a low-frequency, nonoscillatory component of the radiation field that provides a potentially powerful but as yet undetected probe of strong-field gravity. We present the first calculation of gravitational memory from full inspiral--merger--ringdown waveforms in a theory beyond general relativity, focusing on scalar-Gauss-Bonnet gravity as a theoretically well-motivated and numerically accessible extension of GR. Starting from the general memory formulas in Horndeski gravity, we derive explicit spin-weighted spherical-harmonic expressions for the tensor null memory in scalar-Gauss-Bonnet theory and evaluate them on existing numerical-relativity waveforms for both shift-symmetric and dynamically scalarizing binary black hole mergers. We find that the dominant effect is an indirect modification of the tensor memory through changes in the nonlinear merger dynamics, while the direct scalar contribution to the tensor memory remains suppressed by orders of magnitude for the systems considered in this work. For the largest deviations in our dataset, the final memory amplitude differs from the corresponding GR prediction by a few percent and by up to $\sim 4\%$ when compared to the GR template that minimizes the waveform mismatch in a detector-oriented analysis. We further show that including memory increases the mismatch between GR and scalar-Gauss-Bonnet waveforms by more than an order of magnitude, indicating that memory can provide complementary information for testing gravity with third-generation detectors, especially for low-mass binaries.

\end{abstract}

\maketitle


\section{Introduction}


We have now deeply dived into the era of gravitational wave (GW) astronomy, the LIGO-Virgo-Kagra (LVK) collaboration has detected hundreds of GW events from binary mergers~\cite{LIGOScientific:2025slb,LIGOScientific:2025hdt, LIGOScientific:2025yae} and Pulsar Timing Arrays showing convincing evidence of a GW background compatible with inspiraling SMBH~\cite{NANOGrav:2023gor, Reardon:2023gzh, EPTA:2023fyk}. This data is already providing us valuable information about astrophysical population, properties of compact objects, and important tests of our theory of gravity. Next-generation interferometers such as Einstein Telescope~\cite{ET:2025xjr} and Cosmic Explorer~\cite{Reitze:2019iox} on the ground, and LISA~\cite{LISA:2024hlh} and TianQin \cite{TianQin:2020hid} in space, are going to enlarge the visible volume, thus detecting many more events and with greater precision. In particular, they will have the potential to disentangle yet undiscovered effects in the gravitational radiation that could serve as important tests for General Relativity (GR) and potentially point towards a specific type of modified gravity theories. 

Among these effects, in this work we focus on the \textit{gravitational memory effect}, also known as \textit{displacement memory}, namely a persistent offset of the gravitational radiation strain that leaves a long-lasting displacement between two test masses. This effect comes in two forms. The first one was discovered in 1974 by Zel'dovich and Polnarev~\cite{Zeldovich:1974gvh} and developed by Braginskii, Grishchuk~\cite{Braginsky:1985vlg} and Thorne~\cite{Braginsky:1987kwo}. It is related to the emission of unbound matter and is called \textit{linear memory} as it is generated at linear order in perturbation theory. The second one, known as the \textit{nonlinear memory effect}, was predicted in 1991 and is generated by the emission of energy flux towards null infinity~\cite{PhysRevD.45.520,Christodoulou:1991cr,Blanchet:1992br, Wiseman:1991ss, Favata:2009ii}. It was later realized that memory effects have a deep connection both to asymptotic symmetries, encoded in the Bondi-Metzner-Sachs (BMS) group~\cite{Bondi:1962px,Sachs:1962wk}, and to soft theorems \cite{Weinberg_PhysRev.140.B516}. More precisely, memory effects, asymptotic symmetries, and soft theorems form the three corners of the so-called infrared triangle in gauge theories and gravity~\cite{Strominger:2017zoo,Strominger:2013jfa,He:2014laa,Strominger:2014pwa,Pasterski:2015tva}.
More precisely, the dominant displacement memory considered in this work is directly associated with the BMS supertranslation extension of the asymptotic Poincar\'e symmetry, while other subleading persistent observables also exist~\cite{Pasterski:2015tva,Flanagan:2018yzh,Grant:2021hga}. Its origin can be understood as a back-reaction of the emission of energy-momentum from a localized source. In particular, the non-linear memory is associated with the emission of gravitational waves, which in turn generates radiation with a shape markedly different from the dominant oscillatory signal. More specifically, it corresponds to a non-oscillatory, direct-current (DC) component, with most of its support at low frequencies where current GW interferometers are less sensitive. This is one of the main reasons why the memory effect has not yet been detected~\cite{Hubner:2019sly,Hubner:2021amk,Grant:2022bla,Gasparotto:2023fcg,Inchauspe:2024ibs, Cogez:2026frh, Inchauspe:2024ibs, Cogez:2026frh,Zosso:2026czc}.


For the current LIGO–Virgo–KAGRA (LVK) network, a detection of gravitational memory is expected to require stacking $\mathcal{O}(2000)$ events, potentially achievable by the fifth observing run~\cite{Hubner:2019sly, Hubner:2021amk,Grant:2022bla}.
Next-generation detectors, such as the Einstein Telescope and the Cosmic Explorer, are expected to improve the prospects significantly, with predictions of $\sim 7$–$8$ stellar-mass binary black hole (BBH) events per year with detectable memory~\cite{Grant:2022bla}.
In the space-based band, LISA is expected to observe the memory signal from a few to several massive black hole binary mergers, depending on the astrophysical population~\cite{Gasparotto:2023fcg, Inchauspe:2024ibs, Cogez:2026frh,Zosso:2026czc}. Pulsar Timing Arrays are also searching for memory bursts from supermassive black hole binaries without a detection, currently placing upper limits of $\sim 10^{-14}$ on the final amplitude of the memory event~\cite{Tomson:2025ixn}. 

Given these developments and the prospect of a first detection in the near future, a natural question is how gravitational memory can be used to test General Relativity.
To tackle this question, one needs to understand how the memory is modified in other theories. While the oscillatory GW signal from compact binary coalescences (CBCs) has already been studied in several beyond-GR theories, the study of the associated memory remains much less developed. In this work, we focus in particular on modifications of gravity involving an additional non-minimally coupled scalar field, which introduces an extra dynamical degree of freedom within the gravity sector.

The theoretical foundation of our work is provided by
Refs.~\cite{Heisenberg:2023prj,Heisenberg:2024cjk,Zosso:2025ffy}, which derived the gravitational memory for full Horndeski gravity \cite{Horndeski:1974wa,Nicolis:2008in,Deffayet:2009wt,Deffayet:2009mn,Heisenberg:2018vsk,Kobayashi:2019hrl}, the most general gravitational theory with second-order equations of motion possessing a single additional scalar degree of freedom. At the same time, these works provide an understanding of the structure of the memory formula for generic scalar-vector-tensor effective field theories (EFTs). Based on the Isaacson effective-stress-energy-tensor approach, they show how the memory is sourced by all additional radiating degrees of freedom present in a given gravity theory. In that case, the memory in the tensor sector is modified mainly in two ways. First, other radiation channels modify the dominant GW signal, thus resulting in a modified memory. Second, additional radiation channels directly source the memory. 
The concrete case of Horndeski theory incorporates both the case of standard scalar-tensor theory, i.e., Brans-Dicke theory~\cite{Fierz:1956zz,PhysRev.124.925}, as well as the higher-order curvature theory of scalar-Gauss-Bonnet (sGB) gravity \cite{Zwiebach:1985uq,Gross:1986iv,Boulware:1986dr,Moura:2006pz,Nojiri:2005vv,Nojiri:2006je,Pani:2009wy,Pani:2011xm}. For these cases, the explicit connection with asymptotic symmetries have also been demonstrated~\cite{Tahura:2020vsa, Tahura:2021hbk,Hou:2020tnd,Hou:2020wbo, Tahura:2025ebb, Du:2016hww, Koyama:2020vfc,Maibach:2026wpz}. Memory effects have been studied as well in Chern-Simons theory~\cite{Hou:2021oxe,Hou:2021bxz}, in Einstein-Aether theory~\cite{Heisenberg:2025tfh} and Generalized Proca gravity~\cite{Heisenberg:2025roe}. 

However, the above works focus on describing the formula for gravitational memory through its relation to emitted GW energy-momentum and its theoretical implications, but mostly do not provide specific memory waveform models beyond GR that can be searched for in gravitational radiation data. Such explicit results of memory in the tensor and scalar polarizations of binary black hole (BBH) coalescences have been calculated only for the Brans-Dicke theory using a Post-Newtonian expansion~\cite{Lang:2013fna,Tahura:2021hbk, Sennett:2016klh,Bernard:2022noq,Trestini:2024zpi}. This approach can be valid only during the inspiral and omits the merger where the Post-Newtonian expansion is not applicable. On the other hand, the non-linear memory effect of BBHs is theorized to be the strongest during the merger. In addition, the original Brans-Dicke theory is severely constrained by observations \cite{Freire:2024adf} and black holes in this theory cannot carry a nontrivial scalar field. Therefore, the extension of the memory effects to numerical inspiral-merger-ringdown (IMR) waveforms is crucial to answer the question of whether beyond GR effects can be detected by future gravitational wave instruments. In this paper, we make an important step in this direction. Namely, we employ existing full numerical waveforms in the specific beyond GR theories class of sGB gravity and perform the first investigation of the memory effect during the late inspiral and merger. We will work in the assumption of asymptotic flatness and compute the associated gravitational memory using the results of Ref.~\cite{Heisenberg:2023prj} based on the corresponding leading order oscillatory radiation of the BBH events.


While the focus on sGB gravity is driven by the current scarcity of numerical IMR waveforms in theories beyond GR, this theory is also of particular theoretical interest because it occupies a special place among higher-curvature extensions of gravity \cite{Simon:1990PhysRevD41,Yunes:2013dva,Zosso:2024xgy}. From an effective field theory (EFT) perspective, sGB gravity arises naturally as the leading higher-derivative correction to GR within the most general parity-invariant scalar-tensor theory truncated at four derivatives \cite{Weinberg:2008hq}.\footnote{At that order, the action generically contains curvature-squared operators, scalar self-interactions, and mixed scalar-curvature terms. However, many of these are redundant and can be removed by integrations by parts and field redefinitions. After this reduction, a coupling of the form $f(\phi) {\mathcal R}^2_{\rm GB}$, with ${\mathcal R}^2_{\rm GB}$ the Gauss--Bonnet invariant, provides a representative parametrization of the leading nontrivial scalar-curvature interaction.} At the same time, Gauss-Bonnet-dilaton couplings were also motivated by their appearance in the low-energy effective actions of heterotic and bosonic string theory compactifications \cite{Zwiebach:1985uq,Gross:1986iv,Boulware:1986dr,Metsaev:1987zx,Torii:1996yi,Moura:2006pz,Nojiri:2017ncd,Ortega:2024prv}. Within such expansions, the sGB correction is especially appealing because, unlike generic higher-derivative corrections, it belongs to the Horndeski class and yields second-order equations of motion, thereby avoiding the Ostrogradsky instabilities that typically plague higher-derivative theories \cite{Ostrogradsky:1850fid,Woodard:2015zca,deRham:2016wji,Horndeski:1974wa,Nicolis:2008in,Deffayet:2009wt,Deffayet:2009mn,Heisenberg:2018vsk,Kobayashi:2019hrl}.

As such, sGB admits a consistent hyperbolic evolution and well-posed Cauchy problem of its full equations of motion, provided the evolution remains within the regime of validity of the EFT in which the derivative expansion itself remains reliable \cite{Kovacs:2020ywu,Kovacs:2020pns}
. Thus, unlike other higher-curvature theories, sGB gravity need not be treated purely as an order-reduced perturbative correction, while its physically controlled use is still tied to the EFT regime. This makes it a particularly well-motivated framework in which to study strong-field deviations from GR.

From a practical point of view, sGB gravity is especially useful for our purposes because it is, to our knowledge, the only modified theory of gravity for which both scalarized static and rotating black-hole solutions are available \cite{Kanti:1995vq,Torii:1996yi,Pani:2009wy,Kleihaus:2011tg,Sotiriou:2013qea,Doneva:2017bvd,Silva:2017uqg,Antoniou:2017acq,Cunha:2019dwb,Collodel:2019kkx,Berti:2020kgk,Herdeiro:2020wei} and for which fully nonlinear numerical-relativity merger simulations have been achieved with satisfactory accuracy in the weak-coupling regime \cite{East:2020hgw,East:2021bqk,AresteSalo:2022hua,AresteSalo:2023mmd,Corman:2022xqg,Corman:2024vlk,AresteSalo:2025sxc}. In addition, sGB gravity can produce qualitatively new gravitational-wave signatures \cite{Lara:2025kzj,Corman:2025wun,Capuano:2026lhs}. These theoretical motivations, together with the existence of waveform models and numerical simulations, make sGB an especially compelling framework in which to investigate the impact and detectability of gravitational memory beyond GR.

The remainder of this paper is organized as follows. In Sec.~\ref{sec:theoriesbeyondGR}, we review the general framework of gravitational memory and the concept of gravitational polarizations in metric theories of gravity, and summarize the corresponding formulas for Horndeski theories. We then introduce scalar-Gauss-Bonnet gravity in Sec.~\ref{sec:sGBtheory}, where we also derive the concrete null-memory expressions used in our analysis, and summarize the numerical-relativity waveforms employed in this work in Sec.~\ref{sec:numericalwaveforms}. The memory computed from full inspiral--merger--ringdown waveforms is presented in Sec.~\ref{sec:results}, first for shift-symmetric sGB gravity [Sec.~\ref{sec:shiftsym}] and then for dynamically scalarizing binaries [Sec.~\ref{sec:synscal}]. We next investigate the observational impact of these deviations through a detector-oriented mismatch analysis in Sec.~\ref{sec:observability}, before concluding in Sec.~\ref{sec:conclusion}. Technical derivations of the spin-weighted spherical-harmonic expressions for the memory are collected in Appendix~\ref{App:Derivations}, Appendix~\ref{App:BalanceLaws} discusses the total displacement-memory offset from the generalized BMS balance laws, and Appendix~\ref{app:mismatchDS} presents the mismatch results for the dynamical-scalarization case. Throughout this work, we use geometrized units $G=c=1$.

\section{Memory in Theories Beyond GR}\label{sec:theoriesbeyondGR}


\subsection{Metric Theories of Gravity and the Geodesic Deviation Equation}

A broad and physically well-motivated class of theories of gravity is provided by \emph{metric theories of gravity} \cite{Dicke:1964pna,poisson2014gravity,papantonopoulos2014modifications,Will:2018bme,YunesColemanMiller:2021lky}, defined by the assumption that gravity is mediated by a physical spacetime metric to which all matter fields couple universally and minimally. This assumption is equivalent to imposing the Einstein Equivalence Principle (EEP) \cite{Weinberg1972,WaldBook,misner_gravitation_1973,poisson2014gravity,Will:2018bme,carroll2019spacetime}, which is at the root of the formulation of general relativity. It guaranties the local validity of special relativity in freely falling frames and ensures that non-gravitational physics is governed by the same laws as in Minkowski spacetime. As a consequence, freely falling test bodies follow geodesics of a single physical metric, independently of their internal composition or structure. Yet, this assumption does not uniquely determine general relativity, but instead allows for the broader class of metric theories, whose most notable difference from GR is the possibility of admitting additional gravitational dynamical degrees of freedom that may be described in the action by non-minimally coupled fields.

An important implication of the EEP, or equivalently of universal and minimal coupling, is that the relative motion of nearby freely falling observers is governed by a well-defined \emph{geodesic deviation equation}. Crucially, the form of this equation is universal across all metric theories of gravity: it depends only on the spacetime geometry encoded in the physical metric and not on the specific field equations that determine that metric. In particular, the relative acceleration between nearby timelike geodesics separated by a spacial proper distance $s_i$ is determined by the electric-parity components of the Riemann curvature tensor constructed from the physical metric \cite{misner_gravitation_1973,maggiore2008gravitational,carroll2019spacetime} 
\begin{align}
    \ddot{s}_i=-R_{i 0 j 0} s_j \, , \label{eq:geodesicdeviation}
\end{align}
where a dot represents a derivative with respect to proper time of the test masses.

This universality in particular provides a theory-independent kinematical framework for describing the effects of gravitational radiation on detectors. More precisely, in the radiation-zone of an asymptotically flat spacetime, the first order perturbation of the Riemann tensor $\phantom{}_{\myst{(1)}}R_{0i0j}$ is manifestly gauge invariant and is governed by up to six \emph{gravitational polarizations} of the theory, describable in terms of a symmetric spatial polarization tensor $P_{ij}$ satisfying \cite{Flanagan:2005yc,Heisenberg:2024cjk}
\begin{align}
    \phantom{}_{\myst{(1)}}R_{0i0j} = -\frac{1}{2} \ddot P_{ij} \, . \label{eqn:Apolarization}
\end{align}
In particular $P_{ij}$ contains the two transverse-traceless (TT) tensor polarizations $h^{\rm TT}_{ij}$ familiar from GR together with at most four additional polarizations, two vector and two scalar polarizations. The amount of active gravitational polarization modes in $P_{ij}$ is determined by the equations of motion of the theory. Within GR, the Einstein equations dictate that its two dynamical degrees of freedom also excite the corresponding tensor polarizations, while all other components of $P_{ij}$ remain non-dynamical \cite{Flanagan:2005yc}. Similarly, the equations of motion of a specific metric theory beyond GR will determine how many of the six polarization modes in $P_{ij}$ are excited \cite{Heisenberg:2024cjk}.

With this equation at hand, the geodesic deviation equation [Eq.~\eqref{eq:geodesicdeviation}] is easily integrated to give to leading order
\begin{align}\label{eq:integratedGeodesicDeviation}
    \Delta s_i (u) = \frac{1}{2} \Delta P_{ij}(u) s^j (u_0) \,,
\end{align}
where
\begin{equation}
    \Delta s_i (u)\equiv s_i (u)-s_i (u_0)\,,
\end{equation}
for some initial reference proper time $u_0$ before the presence of any gravitational radiation.\footnote{It is convenient to choose a gauge in which $P_{ij}(u_0)=0$.} GWs within the radiative curvature can then be measured as oscillatory changes of the proper distance between freely falling test masses $\Delta s_i (u)$. 

However, asymptotic radiation does not only contain wave perturbations, but also very generically includes a non-wave-like component that leaves a permanent change of the asymptotic background spacetime known as gravitational memory.
Concretely, within the framework described above, the gravitational memory effect can be defined as a permanent change in the relative separation of test masses after the passage of a transient gravitational radiation signal
\begin{equation}\label{eq:MemoryDef}
    \lim_{u\rightarrow\infty}\Delta s(u)\neq 0\,.
\end{equation}
Therefore, the memory effect relies on a non-zero difference before and after the passage of gravitational radiation within one of the components of the polarization matrix 
\begin{equation}\label{eq:Defmemory}
    \lim_{u\rightarrow\infty} \Delta P_{ij}(u)\neq 0\,.
\end{equation}

Importantly, because memory is defined as a net offset in the solution of the geodesic deviation equation, it is a well-defined observable in any metric theory of gravity, irrespective of the detailed form of the gravitational field equations. In this sense, gravitational wave memory is not intrinsically tied to the nonlinear structure of Einstein's equations but is instead a robust consequence of radiative curvature reaching null infinity. What changes from theory to theory is the number of polarization modes that can contribute to the radiative curvature in Eq.~\eqref{eqn:Apolarization} and hence to the memory signal. Very generally, metric theories of gravity may therefore admit \emph{tensor memory}, resulting from a non-zero memory offset within the tensor polarizations, as well as corresponding \emph{vector memory} and \emph{scalar memory}.\footnote{However, note that this nomenclature has nothing to do with the type of the dynamical degrees of freedom which source the memory component. Indeed, as we will explicitly see below, an emission of energy-momentum in the form of scalar radiation sources tensor memory.}

\subsection{Gravitational Memory of Horndeski gravity}


The observation of the general well-definiteness of memory in metric theories of gravity underlies recent systematic analyses of gravitational memory beyond general relativity \cite{Heisenberg:2023prj,Heisenberg:2024cjk,Zosso:2025ffy,Heisenberg:2025tfh,Heisenberg:2025roe}. In particular, it has been shown that for a large class of metric theories satisfying the EEP, the functional form of the memory observable characterized by its final offset within the tensor polarizations of $P_{ij}$ in Eq.~\eqref{eq:Defmemory} is preserved: the permanent displacement measured by idealized detectors within the tensor polarization is always determined by the integrated effect of the energy-momentum content emitted from a localized source. What changes from theory to theory is the set of fields that can contribute to this radiative energy-momentum that source the tensor memory. In particular, also dynamical degrees of freedom that do not directly contribute to the measurable gravitational polarizations influence the gravitational memory.

To render this discussion concrete and precise, we will now turn to the specific and physically relevant example of Horndeski gravity \cite{Horndeski:1974wa,Nicolis:2008in,Deffayet:2009wt,Deffayet:2009mn,Heisenberg:2018vsk,Kobayashi:2019hrl}, the most general scalar–tensor theory with second-order field equations. As such, Horndeski gravity encompasses most of the widely used ghost-free scalar-tensor theories such as Brans-Dicke gravity or scalar-Gauss-Bonnet theory. The Horndeski action employed in this work is written explicitly in Eq.~(5.11) of Ref.~\cite{Zosso:2024xgy} and is characterized by the functionals $\{\bar{G}_i(\Phi,X)\}_{i=2,3,4,5}$, where $X=-1/2(\nabla\Phi)^2$.  

Consider therefore waves radiated from a localized source in asymptotically flat spacetime 
\begin{equation}
    g_{\mu\nu}=\eta_{\mu\nu}+h_{\mu\nu}+\mathcal{O
    }(1/r^2),
\end{equation}
and scalar field expansion
\begin{equation}
    \Phi=\varphi_0+\varphi_1+\mathcal{O
    }(1/r^2),
\end{equation}
with $\varphi_0$ a constant. In Horndeski gravity, the leading order $\mathcal{O}(1/r)$ gravitational wave degrees of freedom consist of two familiar transverse tensor modes 
\begin{equation}
   h^{\rm TT}_{ij}=h_+ e^+_{ij}+h_\times e^\times _{ij}
\end{equation}
and an additional propagating scalar degree of freedom $\varphi$.
Here, a transverse-traceless superscript TT denotes to a projection onto the TT component of the plane-wave solutions via 
\begin{equation}\label{eq:Projectors}
    \perp_{ijab}\equiv \perp_{ia}\perp_{jb}-\frac{1}{2}\perp_{ij}\perp_{ab}\,,
\end{equation}
with
\begin{equation}
\perp_{ij}\equiv \delta_{ij}-n_in_j=u_iu_j+v_iv_j\,.
\end{equation}
The usual source-centered spacial orthonormal basis of the radial outgoing wavefront $\{n_i,u_i,v_i\}$ satisfying 
\[\delta_{ij}=n_in_j+u_iu_j+v_iv_j\] is given by the longitudinal direction $n^i(\Omega)$ and two transverse vectors $u^i(\Omega)$ and $v^i(\Omega)$ with $\Omega=\{\theta,\phi\}$ the spherical angles. Moreover,
\begin{equation}
e^+_{ij}=u_iu_j-v_iv_j\,,\quad e^\times_{ij}=u_iv_j+v_iu_j.
\end{equation}
\newline
\subsubsection{Polarizations of Horndeski gravity}
Based on these dynamical degrees of freedom, the polarization tensor dictating the geodesic deviation in Eq.~\eqref{eq:integratedGeodesicDeviation} is given by \cite{Hou:2017bqj,Heisenberg:2023prj,Heisenberg:2024cjk}
\begin{equation}\label{eq:General Formula ofGWStrain SVH}
P_{ij}=h^{\rm TT}_{ij}-\frac{\bar G_{4,\Phi}}{\bar G_{4}} \varphi_1\, e^{\rm b}_{ij}  +\frac{\bar G_{4,\Phi}}{\bar G_{4}}\left(v^2-1\right)\varphi_1\,e^{\rm l}_{ij}\,,
\end{equation}
where $v$ denotes the asymptotic group velocity of the scalar waves. Thus, depending on the background Horndeski functionals, the theory admits up to two additional gravitational polarizations, a so called breathing polarization characterized by the polarization tensor
\begin{equation}
e^{\rm b}_{ij}\equiv u_iu_j+v_iv_j=\delta_{ij}-n_in_j,
\end{equation}
and a longitudinal polarization
\begin{equation}
 e^{\rm l}_{ij} \equiv n_in_j.
\end{equation}

More precisely, if the scalar field propagates luminally $v=1$ asymptotically, i.e. if the scalar field is massless,
there is no longitudinal polarization, while $\bar G_{4,\Phi}=0$ is equivalent to an absence of any additional polarizations. Note that this condition does not imply the absence of a propagating scalar degree of freedom, but merely the fact that the scalar degree of freedom does not excite an additional polarization within the physical metric, which could directly be detected within the gravitational wave detector response. 

In terms of the gravitational memory effect defined in Eq.~\eqref{eq:MemoryDef}, this implies that, in general, Horndeski admits \emph{tensor memory} within the tensor polarization, as well as \emph{scalar memory}, which can be present both in the breathing or the longitudinal polarizations. In this study, we will focus on the tensor memory within massless Horndeski gravity. The tensor memory formulas for the massive case are derived in Ref.~\cite{Heisenberg:2024cjk}, while the scalar memory will explicitly be discussed in a follow-up paper. In the following, we will distinguish between \emph{null memory} [Sec.~\ref{ssSec:NullMemory}] and \emph{ordinary memory} [Sec.~\ref{ssSec:OrdinaryMemory}]. The notion of null memory generalizes the usual classification of non-linear memory and describes gravitational memory sourced by the emission of energy-momentum that reaches null infinity, thus encompassing both the emission of massless tensor and scalar waves. On the other hand, \emph{ordinary memory}, defined as memory associated with a direct change in supermomentum charge\footnote{The supermomentum charges are the infinite set of charges associated with the BMS supertranslation symmetry at null infinity. They generalize the ordinary Bondi four-momentum by weighting the Bondi mass aspect with an arbitrary function on the celestial two-sphere [see also Appendix.~\ref{App:BalanceLaws}].}, represents a subclass of linear memory and is physically sourced by unbound massive energy-momentum from a localized source.

\

\subsubsection{Null memory in massless Horndeski gravity}\label{ssSec:NullMemory}

Concretely, the tensor null memory in massless Horndeski theory in terms of emitted leading order $\mathcal{O}(1/r)$ gravitational wave degrees of freedom is given by
\cite{Heisenberg:2023prj} 
\begin{widetext}
\begin{align}\label{eq:NonLinDispMemorysGB}
     [\delta h_{ij}]^\text{TT}(u,r,\Omega)=\,\frac{1}{4\pi r}\int_{-\infty}^u d u' \int_{S^2}d^2\Omega'\,r^2 \Bigg\langle |\dot{ h}|^2+\left(3\,\frac{\bar G_{4,\Phi}^2}{\bar G_{4}^2}+\frac{(\bar G_{2,X}-2\,\bar G_{3,\Phi})}{\bar G_4}\right)\,\dot{\varphi}_1^2\Bigg\rangle\,\left[\frac{n'_in'_j}{1-\vec{n}'\cdot\vec{n}(\Omega)}\right]^\text{TT}\,,
\end{align}
\end{widetext}
where we have defined the complex spinweight $s=-2$ tensor scalar
\begin{equation}\label{eq:spinweitght2scalar}
    h\equiv h_+-ih_\times.
\end{equation}
The spacetime averaging $\langle ... \rangle$ over the oscillatory scales of the emitted waves contribution selects out a low-frequency memory observable within the radiation responsible for the defining final offset within the tensor polarizations in Eq.~\eqref{eq:Defmemory} \cite{Heisenberg:2023prj,Zosso:2025ffy,Zosso:2026czc}. Such a spacetime averaging naturally appears in the context of an Isaacson definition of GW energy-momentum \cite{Isaacson_PhysRev.166.1263,Isaacson_PhysRev.166.1272,misner_gravitation_1973,Flanagan:2005yc,maggiore2008gravitational}.
Thus, whenever the scalar degree of freedom is radiative, the associated emission of energy to null infinity
sources an additional contribution to the asymptotic tensor memory. Note that this is independent of the question whether the scalar degree of freedom sources an additional gravitational polarization or not. Even if $\bar G_{4,\Phi}=0$, the scalar degree of freedom contributes to gravitational memory.

Equivalently, this memory formula can also be given explicitly in terms of the modes of a spin-weighted spherical harmonics expansion of the tensor memory 
\begin{align}\label{eq:SW Memory quantity}
   \delta h(u,r,\Omega)&=\frac{1}{2}(e_+^{ij}-i\, e_\times^{ij})[\delta h_{ij}]^\text{TT}\nonumber \\
   &=\sum_{l=2}^\infty\sum_{m=-l}^{l}\, \delta h_{lm}(u,r)\,\,_{\mys{-2}}Y_{lm}(\Omega)\,.
\end{align}
The result reads \cite{Heisenberg:2023prj} 
\begin{equation}\label{NonLinDispMemoryModesSVT}
  \delta h_{lm}=\,\sqrt{\frac{(l-2)!}{(l+2)!}}\int_{S^2}d^2\Omega'\,Y^*_{lm}\int^u d u'\,r\,\bigg\langle |\dot{h}|^2+\rho^2\,\dot{\varphi}_1^2\bigg\rangle\,,
\end{equation}
where we have defined
\begin{equation}\label{eq:DefRho}
 \rho^2\equiv \left(3\,\frac{\bar G_{4,\Phi}^2}{\bar G_{4}^2}+\frac{(\bar G_{2,X}-2\,\bar G_{3,\Phi})}{\bar G_4}\right)\,.
\end{equation}

\subsubsection{Ordinary memory in massless Horndeski gravity}\label{ssSec:OrdinaryMemory}

Note that the memory formulas above exclusively capture the null memory, sourced by unbounded energy-momentum flux that reaches asymptotic null infinity. As mentioned above, there generally also exists an ordinary memory contribution, which is associated with a net change of the supermomentum charge. In the language of the BMS balance laws, gravitational memory is fixed by two distinct ingredients: the integrated null energy flux reaching null infinity, which gives rise to null memory, and the change in the Bondi mass aspect, or more generally in the supermomentum charge, which gives rise to ordinary memory \cite{Zeldovich:1974gvh,Turner:1977gvh,Braginsky:1985vlg,Braginsky:1987gvh,Merritt:2004xa,Gonzalez:2006md,Favata:2008ti,Favata:2008yd,Favata:2010zu,Borchers:2021vyw,Varma:2022pld,DAmbrosio:2024zok} (see Appendix~\ref{App:BalanceLaws} for more details). This ordinary contribution can be generated, for instance, by the emission of unbound massive matter, non-zero remnant kicks or in the context of hyperbolic encounters.

Compared to GR, the change in the supermomentum charge entering the BMS balance laws of massless Horndeski gravity receives an additional correction due to the scalar field \cite{Maibach:2026wpz}.\footnote{To be precise, for full Horndeski theory this formula represents a conjecture. However, it has been confirmed for many subclasses of Horndeski gravity, including Brans-Dicke theory and sGB gravity.}
\begin{equation}
    \Delta\mathcal Q_{\alpha}
    =\int_{S^2} d^2\Omega\,\alpha\left(
    4 \Delta \mathscr M_{\mathrm H}
    -\frac{\bar G_{4,\Phi}}{\bar G_{4}}\mathcal D^2\Delta\varphi_1
    \right)\,,
\end{equation}
where $\alpha=\alpha(\theta,\phi)$ is a function on the 2-sphere $S^2$ that parametrizes the supermomentum charge, and $\mathcal D^2$ is the Laplacian on $S^2$. The generalized Bondi mass aspect in Horndeski gravity reads
\begin{equation}
    \mathscr M_{\mathrm H}
    =\mathscr M_{\mathrm{GR}}
    -\frac{1}{4}\varphi_1\dot\varphi_1\left(
    3\,\frac{\bar G_{4,\Phi}^2}{\bar G_{4}^2}
    +\frac{\bar G_{2,X}-2\,\bar G_{3,\Phi}}{\bar G_4}
    \right)
\end{equation}
with $ \mathscr M_\text{GR}= \mathscr M_\text{GR}(u,\theta,\phi)$ the GR-like Bondi mass aspect that captures the supermomentum charge in the absence of a scalar field.
However, in the context of non-precessing CBCs, the ordinary-memory contribution associated with such changes in supermomentum charge is generally subdominant (see also Appendix~\ref{App:BalanceLaws}), and we therefore focus in the following on the null memory.

\section{Scalar-Gauss-Bonnet gravity }\label{sec:sGBtheory}

As anticipated in the introduction, we will now turn our attention to the particularly interesting theory of scalar-Gauss-Bonnet (sGB) gravity within the Horndeski class. 

\subsection{Definition}
Let's therefore consider the sGB action given by 
\begin{align}\label{eq:ActionsGB}
    S=&\,\frac{1}{16\pi}\int d^4x\sqrt{-g}\left(R-\frac{1}{2}(\nabla\Phi)^2+\lambda\, f(\Phi){\mathcal R}^2_{\rm GB}\right) \notag \\
    &+S_\text{m}[g,\Psi_\text{m}]\,,
\end{align}
where ${\mathcal R}^2_{\rm GB}=R^2-4R_{\mu\nu}R^{\mu\nu}+R_{\mu\nu\rho\sigma}R^{\mu\nu\rho\sigma}$ is the Gauss-Bonnet curvature, $\lambda$ is a coupling constant with dimensions of $[\text{length}]^2$ and $f(\Phi)$ is an arbitrary function of the scalar field $\Phi$. The matter action $S_\text{m}[g,\Psi_\text{m}]$ indicates that the metric $g$ characterizes the physical spacetime of the metric theory which is minimally and universally coupled to matter fields $\Psi_\text{m}$. This theory is indeed part of the full Horndeski class by choosing the Horndeski functionals \cite{Kobayashi:2011nu,Kobayashi:2019hrl}
\begin{equation}\label{eq:CorrespondencesGBHorndeski}
    \begin{split}
        \bar{G}_2&= X+8f^{(4)}( \Phi) X^2(3-\ln X)\,,\\ \bar{G}_3&=4f^{(3)}(\Phi) X(7-3\ln  X)\,,\\
        \bar{G}_4&=1+4f^{(2)}( \Phi) X(2-\ln  X)\,,\\
        \bar{G}_5&=-f^{(1)}( \Phi)\ln  X\,,
    \end{split}
\end{equation}
where $f^{(n)}(\hat \Phi)\equiv\partial^n f/\partial\hat \Phi^n$.

For the function of the scalar field $f(\Phi)$ that couples to the GB curvature we will consider two distinct definite choices: A) a linear coupling, i.e. 
\begin{equation}\label{eq:LinearCoupling}
    f_A(\Phi)=\Phi\,.
\end{equation}
Or B) a quadratic coupling
\begin{equation}\label{eq:GaussianCoupling}
f_B(\Phi)=\frac{1}{2\beta}(1-e^{-\beta\Phi^2})=\frac{1}{2}\Phi^2-\frac{\beta}{4}\Phi^4+\mathcal{O}(\beta^2\Phi^6)\,,
\end{equation}
for a given dimensionless parameter $\beta$. 
The former linear coupling leads to a shift-symmetric theory --given that the action is preserved up to total derivatives under constant shifts in the scalar field-- in which all compact objects possess scalar hair, meaning that they have a non-trivial configuration of the scalar field. In contrast, the latter quadratic coupling function in Eq.~\eqref{eq:GaussianCoupling}, can have both $\Phi=0$ and $\Phi\neq0$ stationary solutions\footnote{Note that the effect of dynamical scalarization can be potentially observed with other simpler types of coupling such as $f=\Phi^2/2$. It leads to unstable scalarized black hole solutions, though, and higher order terms in $\Phi$ are required in order to stabilize them. That is why, for numerical convenience, we have chosen the Gaussian type of coupling, having $f|_{\Phi \rightarrow 0}\cong\Phi^2/2$ leading order expansion for small scalar fields.}. This can lead to dynamical transitions between scalarized and non-scalarized solutions happening not only for isolated objects \cite{Doneva:2017bvd,Silva:2017uqg,Antoniou:2017acq} but also during binary black hole inspiral and merger, such as the so-called phenomenon of dynamical scalarization \cite{Julie:2023ncq,Capuano:2026lhs}. 

\subsection{Memory of scalar-Gauss-Bonnet gravity}\label{sec:memsGBtheory}

As already mentioned in the introduction, the classical sGB gravity is a special theory within the Horndeski class, satisfying $\bar G_{4,\Phi}=0$, hence, only having two tensor gravitational polarizations which can directly be measured within the asymptotic radiation. As a consequence, sGB gravity also only admits tensor memory. Concretely, its dominant null component is given by Eq.~\eqref{NonLinDispMemoryModesSVT} after computing the parameter $\rho$ defined in Eq.~\eqref{eq:DefRho}, which for this specific theory is defined through the choices of functionals in Eq.~\eqref{eq:CorrespondencesGBHorndeski} and is simply given by $\rho^2=1$  \cite{Heisenberg:2023prj}.
Observe that the higher-order sGB term in Eq.~\eqref{eq:ActionsGB} does not directly modify the memory formula since the asymptotic energy-momentum tensor of the waves in sGB is only sensitive to the canonical scalar term. In fact, as understood in \cite{Heisenberg:2023prj}, any term in the action of a local Lorentz preserving EFT involving more than two derivative operators will not modify the tensor memory in an explicit way.

To turn the memory formula in Eq.~\eqref{NonLinDispMemoryModesSVT} into a practical form, we further expand the complex gravitational wave strain $h$ [Eq.~\eqref{eq:spinweitght2scalar}] as well as the scalar field $\varphi_1$ into spin-weighted spherical harmonics (SWSH) just as in Eq.~\eqref{eq:SW Memory quantity}
\begin{align}
   h(u,r,\Omega)&=\sum_{l=2}^\infty\sum_{m=-l}^{l}\, h_{lm}(u,r)\,\,_{\mys{-2}}Y_{lm}(\Omega)\,,\label{eq:SW Memory quantity GR}\\
    \varphi_1(u,r,\Omega)&=\sum_{l=0}^\infty\sum_{m=-l}^{l}\, \varphi_{lm}(u,r)\,\,Y_{lm}(\Omega)\,,\label{eq:SW Memory quantity scalar}
\end{align}
resulting in [see App.~\ref{App:Derivations}]
\begin{align}\label{eq:NonLinDispMemoryModessBG 3}
\delta h_{lm}&=\,r\Bigg[\sum_{l_1l_2\geq 2}\sum_{m_1m_2}\Gamma^{l_1m_1l_2m_2lm}_{-2,2,0} \int_{-\infty}^ud u'\big\langle \dot h_{l_1m_1}\dot h^*_{l_2m_2}\big\rangle\nonumber\\
  &+\sum_{l_3l_4\geq 0}\sum_{m_3m_4}\Gamma^{l_3m_3l_4m_4lm}_{0,0,0} \int_{-\infty}^ud u'\big\langle \dot \varphi_{l_3m_3}\dot \varphi^*_{l_4m_4}\big\rangle\Bigg]\,,
\end{align}
where the factors of $\Gamma^{l'm'm''l''lm}_{s',s'',s}$ are defined in Eq.~\eqref{eq:Gamma}.

For a non-precessing CBC, the spacetime averaging $\langle ...\rangle$ dictates that there is only one main SWSH mode contribution to this time-dependent memory rise, namely the (2,0) mode [see again App.~\ref{App:Derivations}]. In terms of the dominant $(2,2)$ and $(1,1)$ SWSH modes of the primary tensor and scalar waves this memory contribution is given by
\begin{align}\label{eq:Mem20}
     \delta h_{20}(u)=\,&\frac{r}{7}\sqrt{\frac{5}{6\pi}}  \int_{-\infty}^u d u'\big\langle |\dot h_{22}|^2-|\dot \varphi_{22}|^2\big\rangle\nonumber \\
     &-\frac{r}{2\sqrt{30\pi}}  \int_{-\infty}^u d u'\big\langle |\dot \varphi_{11}|^2\big\rangle\,.
\end{align}
The scalar gravitational waves possess a $(1,1)$ dipole radiation which may dominate for sufficient unequal mass ratios of the binaries, while dipole radiation is absent in the equal mass limit \cite{Yagi:2011xp,Corman:2022xqg}.


We want to note here that a quasi-circular binary black hole merger also involves a non-negligible amplitude within the $(0,0)$ monopole radiation. Indeed, while in GR energy-momentum conservation only allows for a quadrupole emission of radiative waves, the absence of scalar charge conservation permits such a leading order change in the scalar monopole. However, Eq.~\eqref{eq:Mem20} indicates that such a monopole radiation will not induce a tensor memory. This is due to the fact that the memory formula in Eq.~\eqref{NonLinDispMemoryModesSVT} is only non-zero for a non-trivial angular dependence of the emitted energy flux. More precisely, its angular integral vanishes for any $l<2$ contribution in a spherical harmonic decomposition of the asymptotic energy flux. Hence, the presence of displacement memory requires a sufficiently spherically asymmetric emission of asymptotic energy. Moreover, the selection rule within the $\Gamma$-factors in Eq.~\eqref{eq:NonLinDispMemoryModessBG 3} would in principle allow for contributions from a hybrid coupling between the large $(0,0)$ mode and the $(2,2)$ mode for instance. However, such an asymmetric contribution contains negligible memory and is thus canceled by the Isaacson averaging. 

The main result of this section is Eq.~\eqref{eq:Mem20}, which will be used to calculate the null memory for the numerical waveforms which capture the fully nonlinear behavior in sGB theory. For the case of quasi-circular binary black hole mergers considered below, this null memory represents the only source of non-negligible memory since, as we explicitly show in App.~\ref{App:BalanceLaws}, the ordinary memory from both black-hole remnant kicks as well as additional scalar terms within the definition of the supermomentum charges are negligible.

\section{Numerical waveforms}\label{sec:numericalwaveforms}
The numerical waveforms employed below are produced during the merger of binary black holes (BBHs) in sGB gravity with the two coupling functions presented in Eqs.~\eqref{eq:LinearCoupling} and \eqref{eq:GaussianCoupling}, i.e. the linear ($A$) and the quadratic ($B$) couplings, and different mass ratios. The simulations are performed using the \texttt{GRFolres} code~\cite{AresteSalo:2023hcp}, an extension of the numerical relativity framework \texttt{GRChombo}~\cite{Andrade:2021rbd} which solves the fully non-linear sGB equations of motion. The waveforms do not constitute a new result, but instead, they are a subset of the waveforms obtained in recent works \cite{AresteSalo:2025sxc,Corman:2025wun,Capuano:2026lhs}. The initial data has been constructed using the GR \texttt{TwoPunctures} spectral solver~\cite{Ansorg:2004ds} integrated into \texttt{GRChombo}, which provides binary puncture data of Bowen-York type. For both types of coupling, these are not constraint-violating initial data. 

In the shift-symmetric linear coupling case, the black holes are always endowed with a scalar field, and thus the GR initial data do not represent a quasi-stationary solution of inspiraling BBHs in sGB gravity. Instead, the scalar field would develop rapidly as GR initial data is evolved in time. In some cases, this can lead to a large increase in the eccentricity which will spoil the comparison with the low-eccentricity GR simulations. To cure that, an eccentricity reduction can be performed,  that is the case for our high-coupling where $\lambda/m_2^2=0.1414$ where $m_2$ is the smaller mass in the close to equal mass binary simulation employed from \cite{Corman:2025wun}. Due to the lower coupling, the unequal mass ratio simulations from \cite{AresteSalo:2025sxc} (with $\lambda/m_2^2=0.106$) do not exhibit a large eccentricity increase. 

For the quadratic coupling, both GR and hairy black hole solutions can co-exist. This leads to the possibility for spontaneous scalarization of black holes \cite{Doneva:2017bvd,Silva:2017uqg}. Namely, for small enough masses the GR black holes lose stability and instead they transition to a new scalarized state. When put in a binary, another interesting effect called dynamical scalarization can be observed. This happens when the two black holes are close to but below the point of spontaneous scalarization when isolated. However, if put in a binary, a scalar field can develop when the distance between the two black holes decreases below a certain threshold (the effect resembles well the dynamical scalarization of neutron stars in pure scalar-tensor theories \cite{Barausse:2012da,Shibata:2013pra}). It is clear then that the GR initial data are the true quasi-stationary solution when the two black holes are far away from each other.


The most stringent observational bound on shift-symmetric sGB gravity comes from comparing the black hole-neutron star GW signal GW230529 to
Post-Newtonian results for EsGB and give a constraint of $\sqrt{\lambda}\lesssim 0.75$ km \cite{Sanger:2024axs}.  If one assumes the smallest BH observed so far ($3.6M_{\odot}$), then the highest coupling value we have considered ($\lambda/m_2^2=0.1414$)
corresponds to $\sqrt{\lambda}\sim 2$ km, i.e. larger by a factor of $\sim2$ than the current bounds we have on this theory. Nevertheless, it is still within the same order of magnitude, in addition to the fact that these bounds rely on the assumption that PN theory is valid up to merger, which may not be valid as shown in \cite{Corman:2025wun}. On the other hand, quadratic type coupling are less constrained. However, one can still apply current observational limits for pulsars as studied in \cite{Danchev:2021tew,Yordanov:2024lfk}. The latter yield slightly different results for different equations of state, ranging from $\sqrt{\lambda}\leq 10.90$ km to $\sqrt{\lambda}\leq 25.44$ km. However, slightly higher values of $\lambda$ are also allowed, as pointed out in \cite{Wong:2022wni}, where it was shown that strongly disfavored values start at $\sqrt{\lambda}\geq 41.28$ km.
For the waveforms we considered, an equal-mass system with a coupling constant of $\lambda/m_2^2=0.703$ from~\cite{Capuano:2026lhs}, this constraint corresponds to a maximum total mass of $\sim 66 M_{\odot}$. 

\begin{figure*}[t]
    \centering
    \includegraphics[width=1\textwidth]{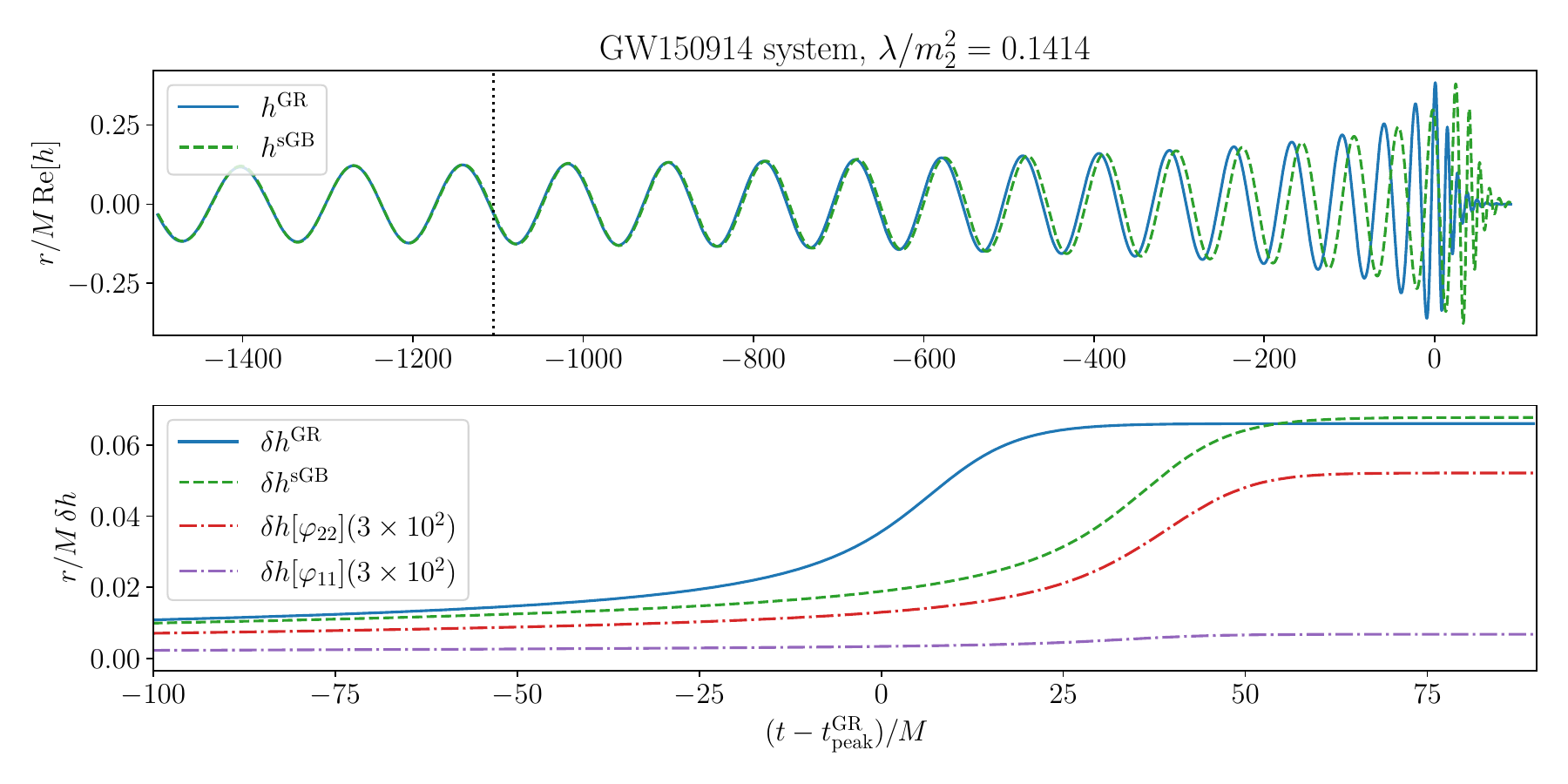}
    \caption{ Gravitational waveforms for a near-equal-mass binary in GR and shift-symmetric sGB gravity with $\lambda/m_2^2=0.1414$ and mass ratio $q=1.221$ (GW150914-like). The final memory amplitude is $0.0634$ in sGB and $0.0619$ in GR at $r=150M$. As the scalar-induced memory is negative, we show its absolute value. The vertical dotted line indicates the end of the alignment interval (see Ref.~\cite{Corman:2025wun}). }
    \label{fig:shiftmain}
\end{figure*}

\section{Results for the tensor and scalar memory}\label{sec:results}

In this section we compute the nonlinear gravitational-wave memory from full inspiral--merger--ringdown waveforms obtained in Refs.~\cite{AresteSalo:2025sxc,Corman:2025wun,Capuano:2026lhs}. In particular, we evaluate the $(2,0)$ mode sourced by both the tensor and scalar strains according to Eq.~\eqref{eq:Mem20}. This approach captures the dominant contribution to the memory arising from the merger phase, where most of the energy in both tensor and scalar radiation is emitted, and which is therefore most relevant for observations.

For each simulation, we compare results obtained in sGB gravity ($\lambda\neq 0$) with the corresponding GR case ($\lambda=0$). The final memory values reported here are computed including only the $(2,2)$ mode, as in Eq.~\eqref{eq:Mem20}, which provides the dominant contribution\footnote{Note that Eq.~\eqref{eq:Mem20} uses the following relation $h_{2,-2}=h^*_{2,2}$, so even if in the main text we refer just to the $(2,2)$ mode, we are actually using both $(2,\pm2)$.}. We have verified that these values are in good agreement with those obtained from the waveform model \texttt{NRHybSur3dq8}~\cite{NRHybSur3dq8} over the same time interval.

Including higher modes increases the total memory by at most $\lesssim 10\%$, indicating that our estimates are conservative. We do not expect the relative difference between the GR and sGB memory to be significantly affected by higher-mode contributions. Similarly, the final value of the memory depends on the total duration of the signal; since our calculation is based on relatively short numerical-relativity waveforms, our estimates are again conservative. However, this limitation is not expected to impact our conclusions, as scalar effects are subdominant during the early inspiral for the considered waveforms.

In Sec.~\ref{sec:shiftsym} we consider the shift-symmetric case [Eq.~\eqref{eq:LinearCoupling}] for near-equal masses and varying mass ratios, while in Sec.~\ref{sec:synscal} we analyze the dynamical scalarization scenario [Eq.~\eqref{eq:GaussianCoupling}].

\subsection{Memory in shift-symmetric sGB gravity}\label{sec:shiftsym}

\subsubsection{GW150914-like event, high coupling}\label{sec:GW150914sys}

\begin{figure}
    \centering
    \includegraphics[width=1\linewidth]{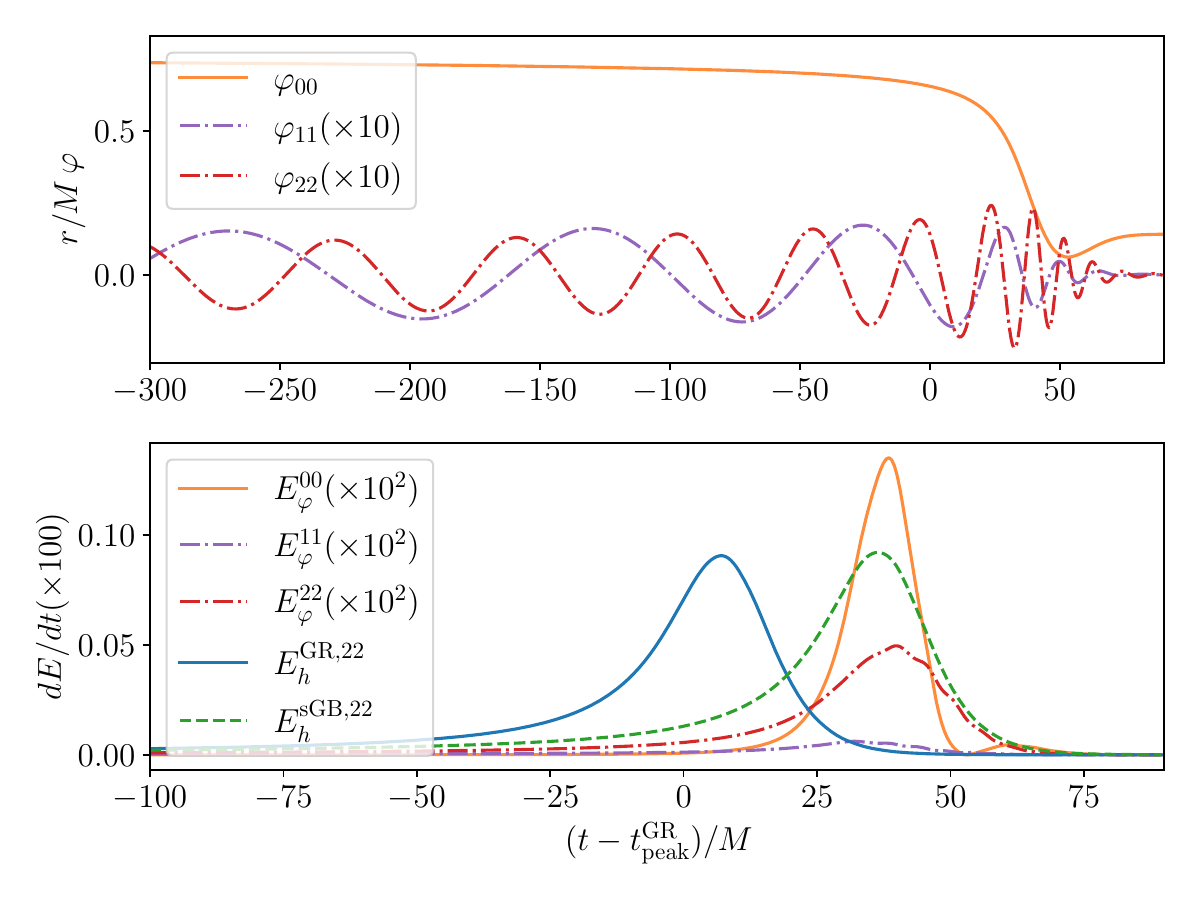}
    \caption{Scalar field and energy fluxes for the same system as in Fig.~\ref{fig:shiftmain}.}
    \label{fig:scalarmain}
\end{figure}

The first waveforms considered are taken from \cite{Corman:2025wun} and correspond to a binary with parameters consistent with those of the first gravitational-wave event detected, GW150914~\cite{LIGOScientific:2016vlm}. The spins are set to zero and the mass ratio is $q = 1.221$, corresponding to individual masses $m_1 = 0.5497M$ and $m_2 = 0.4502M$, where $M = m_1 + m_2$. The GB coupling, normalized to the smaller mass in the system, is $\lambda/m_2^2 = 0.1414$. The dominant $(2,2)$ mode of the tensor waveforms for GR (green) and sGB (dashed blue) is shown in the upper panel of Fig.~\ref{fig:shiftmain}, while the scalar waveforms are shown in Fig.~\ref{fig:scalarmain}. We display the waveforms extracted at radius $r=150M$ in order to avoid potential errors associated with extrapolation to infinity (see discussion below).

In Fig.~\ref{fig:shiftmain}, we adopt the optimal alignment used in \cite{Julie:2024fwy,Corman:2025wun}, which consists in minimizing the difference in frequency and phase over a given time window between the first $400M$ times of the waveform. This alignment shows that, for this system and coupling, the merger in sGB is delayed relative to GR, contrary to the naive intuition that in the presence of an additional channel of energy emission, namely the scalar field, the inspiral is faster. 
In the strong-field regime the modified conservative dynamics in sGB appear to slightly delay the merger and produce a somewhat stronger signal, with a slightly larger amplitude and higher frequency near coalescence. The difference in the peak amplitude of the two waveforms is small: we find $\text{max}\{h^{\rm GR}\}=0.384$ and $\text{max}\{h^{\rm sGB}\}=0.387$, corresponding to a relative difference of less than $1\%$.

As shown in Fig.~\ref{fig:scalarmain}, the dominant scalar radiation is in the $\varphi_{00}$ mode, 
while higher multipoles are suppressed by more than one order of magnitude relative to it. This follows due to the non-trivial scalar charge of the black holes, which yields the following asymptotic behavior of the scalar field: $\Phi\sim \frac{\lambda}{r}(\frac{1}{m_1}+\frac{1}{m_2})$ during the inspiral and $\Phi\sim\frac{\lambda}{rM}$ in the post-merger phase~\cite{AresteSalo:2025sxc}. This explains as well the change of value of $\varphi_{00}$ after merger. 

The bottom panel of Fig.~\ref{fig:scalarmain} shows the flux of energy radiated in the different channels, specifically the $(2,2)$ tensor mode and the $(0,0)$ scalar mode. As one can see, despite the fact that the latter is the dominant scalar mode, it is two orders of magnitude smaller than the tensor one. Higher scalar modes are even smaller. The energy flux radiated in a given mode $(l,m)$ for a waveform $\kappa$, being either tensor $h$ or scalar $\varphi$,  is defined as 
\begin{equation}
\frac{d E^{(l, m)}_\kappa}{d t}=\frac{|\dot \kappa^{(l, m)}|^2}{16\pi}.
\end{equation}
The difference in the energy flux—relevant for understanding the difference in the final memory—is approximately $1.6\%$ higher in sGB compared to GR, while the total energy radiated in the $(0,0)$ scalar mode is $\sim 10^{-4}M$. 

We now turn to the final memory computed from the $(2,2)$ tensor waveforms in GR (blue) and sGB (dashed green), shown in the bottom panel of Fig.~\ref{fig:shiftmain}, together with the memory generated by the $(2,2)$ and $(1,1)$ scalar waveforms. 
In this plot we keep the alignment of the oscillatory waveforms for reference; however, if the waveforms are instead aligned at their respective peak, the main deviation appears as a small difference in the final amplitude, with the memory in sGB reaching a slightly larger value. This is consistent with the fact that the GW radiation is slightly stronger in the sGB case. The peak value of the memory in GR is $0.066$, while in sGB we find $0.068$, corresponding to a difference of order $2.5\%$. Interestingly, the difference in the peak value of the memory strain is amplified relative to that of the oscillatory waveform. This is a consequence of the hereditary nature of the memory, which involves an integration over the entire waveform evolution. As a result, small numerical errors in the oscillatory waveform—for instance those arising from extrapolation to infinity—can propagate and be amplified in the memory calculation. We verified that the result remains unchanged when computed using waveforms extracted at different radii, as shown in Table~\ref{tab:radius_gb_gr}. The variations between different extraction radii are smaller than $\mathcal{O}(0.1\%)$, confirming that the $\sim2\%$ difference between GR and sGB is robust. 
Note that, in computing the memory, we preserve the alignment of the oscillatory waveforms. In particular, the integration is performed starting from a common reference time within the interval used to minimize the frequency mismatch. As a consequence, the integration window in the GR case is slightly shorter, since the binary merges earlier\footnote{For the computation of the memory we use the same time interval around merger for both waveforms, $t_{\rm min}=-1500M$ to $t_{\rm max}=90M$, where $t_0=0$ corresponds to the peak of the GR waveform.}.
We have verified that adopting instead an alignment at the respective peak amplitudes and integrating over the same time interval leads to only a small difference, at the level of $\mathcal{O}(0.1\%)$, as reported in Table~\ref{tab:radius_gb_gr}.

Moreover, the final value of the memory depends on the full integration interval. We find that using \texttt{NRHybSur3dq8} waveforms, passing from $t_{\rm in}=-1000M$ to $t_{\rm in}=-10000M$, increases the amplitude of the memory by roughly 10\%. As mentioned earlier, this would have a negligible impact on the final difference of the memory between GR and sGB waveforms and has no impact on the observable signature of memory which is generally confined to the buildup within a window of $\sim 40 M$ around merger of quasicircular binaries \cite{Zosso:2026czc}. 

We stress that the final difference in memory amplitudes arises primarily from sGB-mediated modifications to the nonlinear dynamics of the gravitational-wave emission, while the direct back-reaction of energy-momentum emission in the scalar field remains negligible. Concretely, the scalar-emission induced memory contributions are suppressed by two and three orders of magnitude for the $\varphi_{22}$ and $\varphi_{11}$ modes, respectively. In the system considered here, which is close to equal mass, the $(2,2)$ scalar mode dominates over the $(1,1)$ mode. However, this hierarchy changes as the mass ratio increases, as discussed in the next section. Notably, the dominant scalar radiation emission in the $(0,0)$ mode does not directly generate memory because of its spherical symmetry, as discussed in Sec.~\ref{sec:memsGBtheory}. 



\begin{table*}[t!]
\centering
\begin{tabular}{lccccc}
\hline
 & \text{aligned at max (inf)} & \text{aligned at freq (inf)} & $R=100M$ & $R=150M$ & $R=200M$  \\
\hline
sGB & \texttt{0.0634} & \texttt{0.0635} & \texttt{0.068} & \texttt{0.067} & \texttt{0.0648}\\
GR            & \texttt{0.0619} & \texttt{0.0619} & \texttt{0.066} & \texttt{0.066} & \texttt{0.0631} \\

Ratio\%            & \texttt{2.36} & \texttt{2.53} & \texttt{2.54} & \texttt{2.54} & \texttt{2.52} \\
\hline
\end{tabular}
\caption{Final values of the memory computed from waveforms extracted at different radii in the sGB and GR cases. ``Aligned at max (inf)'' indicates that the waveforms are aligned at the peak of the oscillatory signal, using waveforms extrapolated to infinity. ``Aligned at freq (inf)'' instead refers to the alignment used in Fig.~\ref{fig:shiftmain}, where the sGB and GR waveforms are matched at the same common frequency within a time interval prior to merger and with extrapolation to infinity. Waveforms extracted at finite radius are kept with this latter alignment. }
\label{tab:radius_gb_gr}
\end{table*}
\begin{figure*}
    \centering
    \includegraphics[width=1\linewidth]{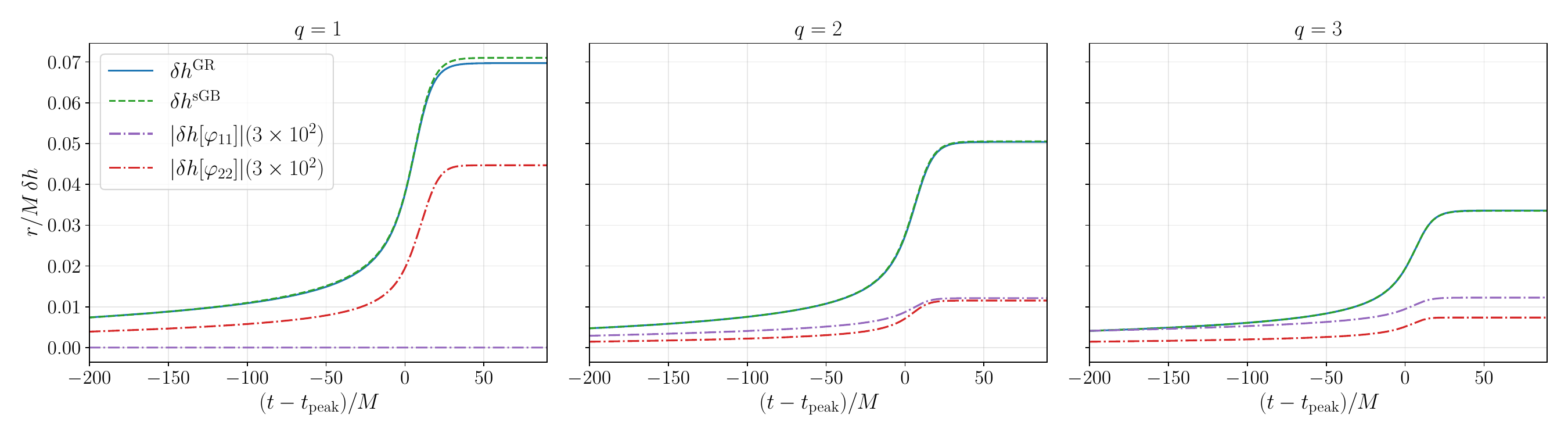}
    \caption{Tensor memory for different mass ratios at fixed coupling $\lambda/m_2^2=0.106$ ($m_2$ is the smaller black hole mass). Waveforms are extracted at $r=150M$ and aligned at the peak of the oscillatory waveform. }
    \label{fig:massratios}
\end{figure*}

\subsubsection{Unequal mass binary}
We now consider waveforms for unequal mass ratios \(q\) presented in \cite{AresteSalo:2025sxc}, corresponding to \(q=1,2,3\). In that work, waveforms are computed for the same normalized coupling \(\lambda/m_2^2\), where \(m_2\) denotes the mass of the smaller black hole. 
Similarly to Fig.~\ref{fig:scalarmain}, we show in Fig.~\ref{fig:massratios} the tensor memory for GR (\(\lambda=0\)) and sGB, together with the contributions obtained from the tensor and scalar waveforms\footnote{In computing the memory we discard the initial portion of the simulations, approximately the first \(200M\), which contains junk radiation.}. For reference, the final values of the memory in the different cases are reported in Table~\ref{tab:memory_unequal}.
In these values the initial memory is set to zero, so the accumulation during the early inspiral— which is not captured in our simulations— is neglected. 

From Fig.~\ref{fig:massratios} we observe that, for the tensor contribution, both the final memory amplitude and the relative difference between GR and sGB decrease as the mass ratio increases. This behavior is expected, since the total gravitational-wave emission decreases with increasing \(q\), which follows from the reduction of the total radiated GW energy as the mass ratio increases~\cite{Islam:2021old}. The memory generated by the scalar radiation in sGB exhibits a different dependence on the mass ratio. In particular, the contribution from the \(\varphi_{22}\) mode decreases rapidly with \(q\) and becomes subdominant already for \(q=2\) compared to the dipole contribution from the \(\varphi_{11}\) mode. Interestingly, while the tensor memory from the \(h_{22}\) mode decreases between \(q=2\) and \(q=3\), the dipole-induced memory remains nearly constant. This behavior is consistent with the enhancement of scalar dipole radiation as the mass ratio departs from unity, since it is sourced by the difference in the scalar charges of the two black holes. However, the memory generated from the scalar sector remains around two orders of magnitude below the tensor memory for these waveforms, thus unobservable for any practical purposes. 

\begin{table}[t!]
\centering
\caption{Finite differences in the memory values for different mass ratios, computed from the waveforms shown in Fig.~\ref{fig:massratios}. The relative difference is defined as $(\delta h^{\rm GB}-\delta h^{\rm GR})/\delta h^{\rm GR}$. }
\label{tab:memory_unequal}
\begin{tabular}{lccc}
\hline
 & $\Delta(\delta h^{\rm GR})$ & $\Delta(\delta h^{\rm GB})$ & \text{Rel difference [\%]} \\
\hline
$q=1$ & \texttt{0.0697} & \texttt{0.0710} & \texttt{1.24} \\
$q=2$            & \texttt{0.0504} & \texttt{0.0505} & \texttt{0.27}\\

$q=3$           & \texttt{0.0335} & \texttt{0.0335} & \texttt{$<0.1$}\\ 
\hline
\end{tabular}
\end{table}

\subsection{Memory for dynamically scalarized black holes}\label{sec:synscal}
\begin{figure*}
    \centering
    \includegraphics[width=1\textwidth]{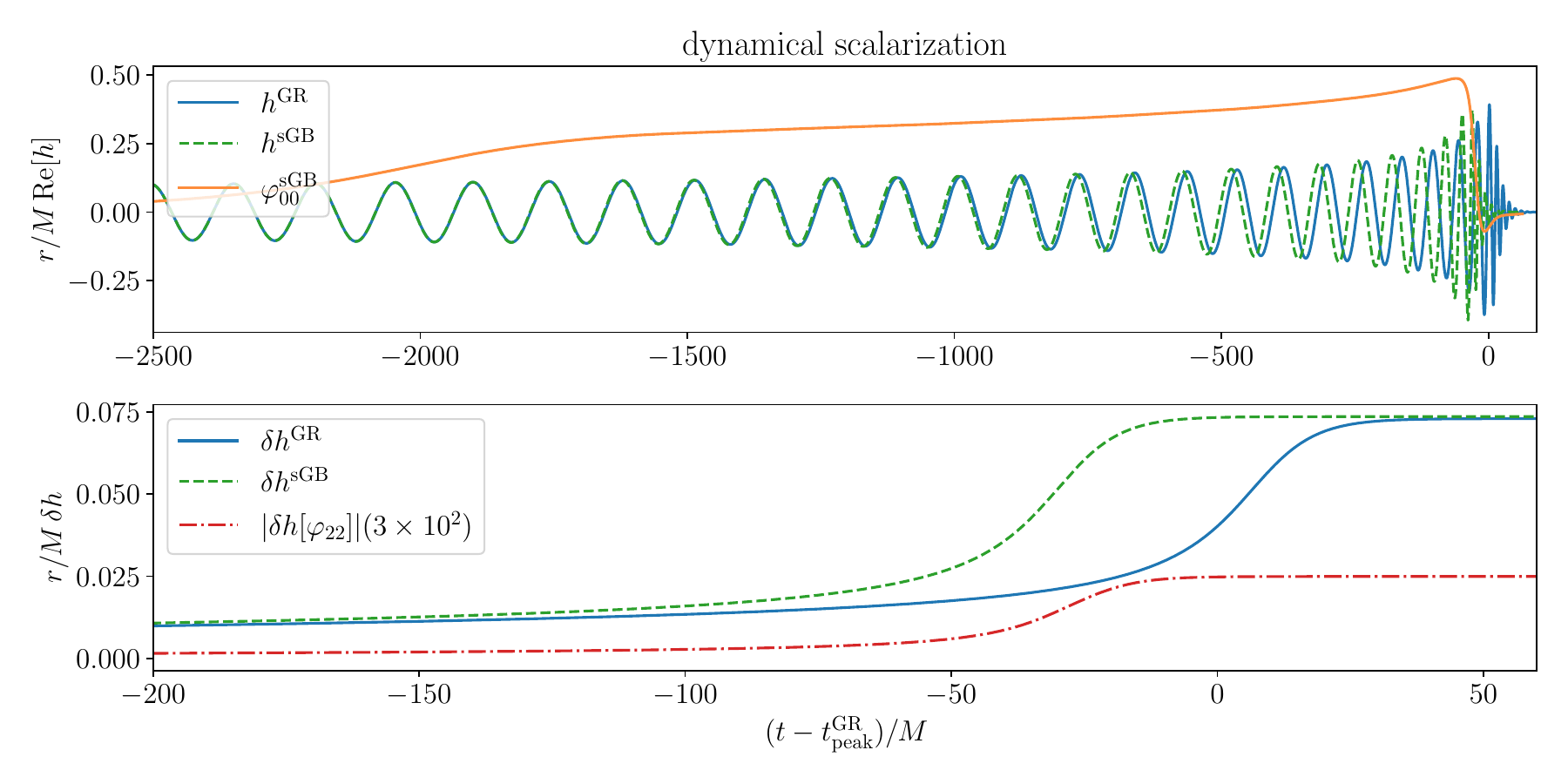}
    \caption{Equal-mass binary in GR and sGB gravity exhibiting dynamical scalarization with quadratic coupling parameters $\lambda/m_2^2=0.703$ and $\beta=4$. We set $t=0$ at the peak of the GR waveform, which occurs later than in the dynamically scalarized case. The final memory amplitude is $0.0729$ in GR and $0.0735$ in sGB.}
    \label{fig:DSwaveforms}
\end{figure*}
\begin{figure*}
    \centering
    \includegraphics[width=0.47\linewidth]{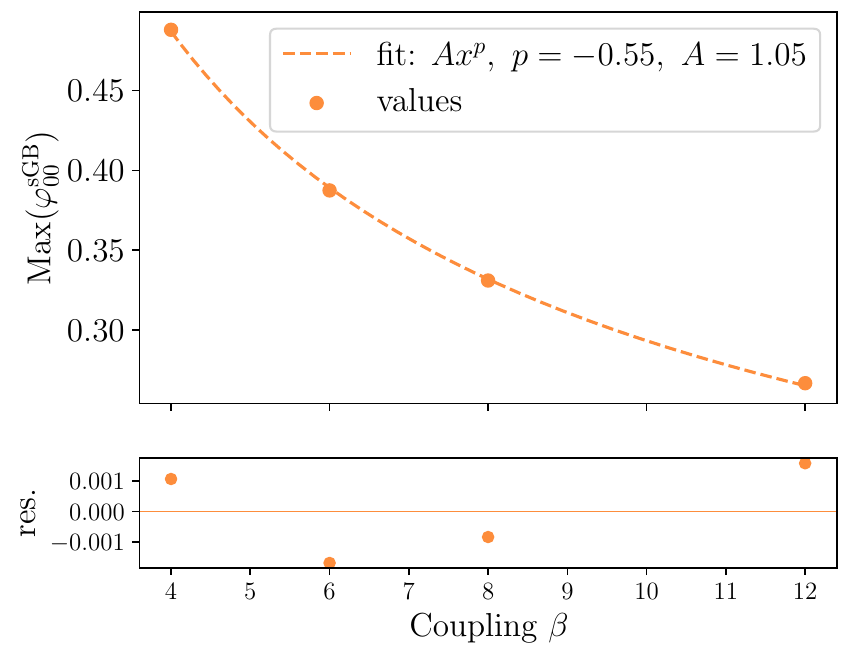}
    \hspace{0.03\linewidth}
    \includegraphics[width=0.47\linewidth]{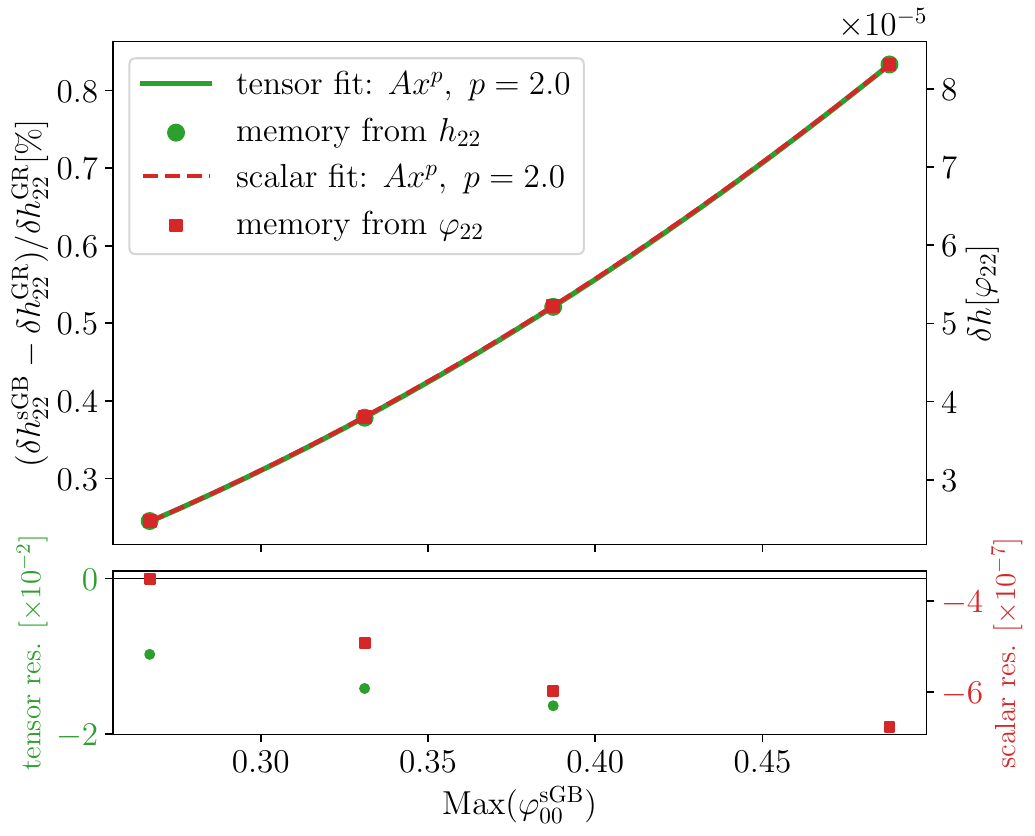}
    \caption{\textit{Left:} Maximum value of the scalar charge $\varphi_{00}$ for the different simulations, corresponding to different values of the coupling parameter $\beta$.
    \textit{Right:} Relative difference (in percent, left y-axis) in the tensor memory sourced by the $h_{22}$ mode, together with the final amplitude of the memory generated by $\varphi_{22}$ (right y-axis), shown as a function of the scalar charge. Both quantities exhibit the same scaling with the scalar charge. Note the different scales used for the two y-axes.}
    \label{fig:DSfits}
\end{figure*}

We now consider the case of a  quadratic coupling between the scalar field and the sGB term in Eq.~\eqref{eq:GaussianCoupling}. We focus on the regime in which the combination of the coupling parameters $\lambda$ and $\beta$ leads to dynamical scalarization. The waveforms employed in this analysis are taken from Ref.~\cite{Capuano:2026lhs}.

We consider equal-mass binaries and fix the quadratic coupling to $\lambda/m_2^2 = 0.703$ (or $\lambda=0$ in the GR case), while varying the dimensionless parameter in the exponent $\beta = \{4,6,8,12\}$.
The onset of scalarization is controlled by $\lambda$, which sets the characteristic radius at which the scalar field becomes unstable, whereas $\beta$ governs the nonlinear dynamics and determines the maximum amplitude reached by the scalar field during the merger before it rapidly decays back to zero since the final post-merger black hole can not sustain scalar hair.

The peak value of the $\varphi_{00}$ mode, which we identify as the scalar charge, scales approximately as $\max\{\varphi_{00}\} \propto \beta^{-1/2}$, as shown in the left panel of Fig.~\ref{fig:DSfits}. Since the modification to the gravitational waveform with respect to GR is driven by the scalar charge, we illustrate in Fig.~\ref{fig:DSwaveforms} only the case $\beta=4$. In this configuration, the final value of the memory is $0.0735$ in the sGB theory and $0.0729$ in GR, corresponding to a relative difference of $\sim 1\%$~\footnote{Note that the slightly larger values found for the memory here, compared to Fig.~\ref{fig:shiftmain}, arise from the different mass ratio ($q=1$ in this case) and from the slightly longer waveforms, with $t_{\rm in} \sim -2700M$.}. For this choice of parameters, the effect is slightly smaller than in the configurations discussed in Fig.~\ref{fig:shiftmain}.

We then use the full set of simulated waveforms to investigate how the memory generated in both the tensor ($h_{22}$) and scalar ($\varphi_{22}$) channels depends on the scalar charge. The results are shown in the right panel of Fig.~\ref{fig:DSfits}. We find that both the difference between sGB and GR tensor memory and the scalar memory scale quadratically with the scalar charge with remarkable accuracy. 

This behavior is nontrivial for two reasons. First, the correction to the tensor memory arises from the effect of the scalar charge on the non-linear dynamics of the system, which primarily modifies the oscillatory gravitational-wave signal. Second, the scalar memory originates from higher multipoles of the scalar field (scaling as $\propto \varphi_{00}$), which contribute subdominantly to the tensor sector. However, the time integration that defines the memory preserves the quadratic scaling of the underlying flux ($\propto \varphi_{00}^2$).  
Despite these different physical origins, both contributions exhibit the same quadratic dependence on the scalar charge. 

In the shift-symmetric case considered above, the scalar charge scales linearly with the coupling, $\propto \lambda$. By analogy with the behavior observed in dynamical scalarization, we therefore expect the relative difference in the memory to scale quadratically with the coupling. We thus conjecture that this quadratic scaling holds in the shift-symmetric case as well, with
\begin{equation}
\frac{\delta h^{\rm GB} - \delta h^{\rm GR}}{\delta h^{\rm GR}} \sim \lambda^2 \, 
\end{equation}
and we have checked this for the system discussed in \ref{sec:shiftsym}.
However, due to the limited number of available simulations, we have not tested this behavior across a broader range of coupling values.


Finally, we emphasize that these results are based on a limited set of simulations and should be confirmed with a more extensive parameter study. Nevertheless, the comparison between sGB and GR waveforms should be robust, as all simulations are performed within the same numerical setup and processed in the same way.

\section{Observability}\label{sec:observability}
So far, we have compared numerical waveforms obtained with and without the sGB coupling, i.e.\ $\lambda \neq 0$ and $\lambda = 0$, for systems with the same total initial ADM mass. As discussed in detail in \cite{Corman:2025wun}, the scalarization of the BHs can potentially change this quantity, but it was verified that this difference was not significant enough, which was also the case in the dynamical scalarization studied in \cite{Capuano:2026lhs}. This should not be confused with the total ADM mass of the remnant, which can be significantly different because of the larger release of energy happening in sGB.  

In this section, we move to a detector-oriented comparison between GR and non-GR waveforms. In a realistic data-analysis context, a non-GR signal would be recovered using GR templates, and differences between the two may be partially degenerate with variations of the intrinsic parameters of the template. For instance, Ref.~\cite{Julie:2024fwy} finds a clear degeneracy between the sGB coupling and the chirp mass, suggesting that larger values of the coupling can mimic systems with smaller chirp mass.
This raises the following questions: which parameter variations minimize the difference between GR and non-GR waveforms? Under which conditions does this difference become observable? What is the corresponding impact on the gravitational-wave memory, and can the latter help break such degeneracies?

In the absence of a full parameter-space exploration, we restrict our analysis to variations of the total mass of the system. This choice is motivated by the fact that mass rescalings induce a simple transformation of the waveform, corresponding to a rescaling of both the time and amplitude.

To quantify the distinguishability between two waveforms, we employ the standard noise-weighted inner product used in gravitational-wave data analysis. Given two signals $h_1(t)$ and $h_2(t)$, their overlap is defined in the frequency domain as
\begin{equation}
\langle h_1 | h_2 \rangle = 4\,\mathrm{Re} \int_{f_{\min}}^{f_{\max}} \frac{\tilde{h}_1(f)\,\tilde{h}_2^*(f)}{S_n(f)} \, df \, ,
\end{equation}
where $\tilde{h}(f)$ denotes the Fourier transform of the waveform and $S_n(f)$ is the one-sided noise power spectral density of the detector and $f_{\min}, f_{\max}$ correspond to the minimum and maximum frequencies of the detector band. The normalized overlap is then
\begin{equation}
\mathcal{O}(h_1,h_2) = \frac{\langle h_1 | h_2 \rangle}{\sqrt{\langle h_1 | h_1 \rangle \langle h_2 | h_2 \rangle}} \, ,
\end{equation}
which takes values between $0$ and $1$ and is maximized over relative time and phase shifts between the signals.

The mismatch is defined as
\begin{equation}
\mathcal{M} = 1 - \max_{\Delta t,\,\Delta \phi} \mathcal{O}(h_1,h_2) \, ,
\end{equation}
and quantifies the fractional loss in signal-to-noise ratio due to modeling differences. This quantity is commonly used to assess waveform distinguishability. In particular, two waveforms are considered indistinguishable according to the indistinguishability criterion~\cite{Flanagan:1997kp,Purrer:2019jcp} if
\begin{equation}\label{eq:distcriter}
    \mathcal{M} \leq \frac{D}{2\,\mathrm{SNR}^2} \,, 
    \qquad \mathrm{SNR}^2 = \langle h_1 | h_1 \rangle \, ,
\end{equation}
where $D$ is the number of intrinsic parameters of the waveform model.

We begin by considering the waveforms discussed in Sec.~\ref{sec:GW150914sys}, which exhibit the largest deviation in the memory signal. We rescale the time and strain to correspond to a binary with total mass $M=20\,M_\odot$, such that the signal is observable with high SNR by the Einstein Telescope. This 
mass lies in the region of the BBH population inferred by the LVK collaboration where the distribution peaks and the associated uncertainties are smallest~\cite{KAGRA:2021duu,LIGOScientific:2025pvj}, which makes our results subject to relatively smaller astrophysical uncertainties. 

\begin{figure}
    \centering
    \includegraphics[width=\linewidth]{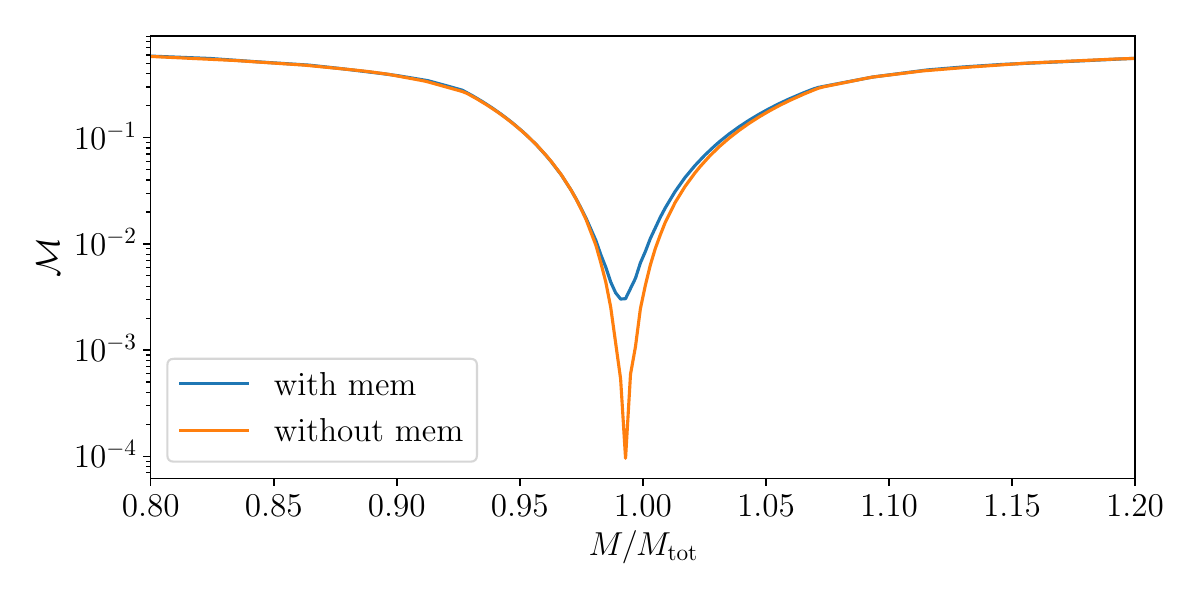}
    \caption{Mismatch between GR and non-GR waveforms for the system in Fig.~\ref{fig:shiftmain}, computed for ET with $M=20\,M_{\odot}$, $d=420\,\mathrm{Mpc}$, and $\iota=90^\circ$. The corresponding signal-to-noise ratio is $\mathrm{SNR}\sim 82$. }
    \label{fig:mismatch_simple}
\end{figure}

We then compute the mismatch as a function of the total mass of the GR template, in order to identify the value that better recovers the non-GR waveform and to quantify the corresponding difference in the final value of the memory.
The resulting mismatch is shown in Fig.~\ref{fig:mismatch_simple}, where we consider an edge-on configuration, which maximizes the memory contribution. We find that including the memory significantly impacts the mismatch, increasing its minimum value by more than an order of magnitude, from $\mathcal{M}\simeq 10^{-4}$ without memory to $\mathcal{M}\simeq 3\times 10^{-3}$ when memory is included. These values are comparable for smaller total masses, for which the memory contribution has a slightly larger impact, as it falls within the most sensitive region of the detector band. Similar analysis is presented for the dynamical scalarization case in Appendix~\ref{app:mismatchDS}; however, in that case the deviation from GR is smaller.

These results should be regarded as conservative estimates, since our waveforms do not include the early inspiral phase, which would both increase the total SNR and provide additional information to break parameter degeneracies. For $M=20\,M_\odot$, the starting frequency of the waveform corresponds to $f_{\rm min}^{\rm ins}\sim 80\,\mathrm{Hz}$, while the memory contribution extends the signal to lower frequencies. In our analysis, we adopt the sensitivity curve of Einstein telescope, the ET-D in~\cite{Hild:2010id}, with frequency range $f_{\rm min}=5\,\mathrm{Hz}$ and $f_{\rm max}=5\times10^3\,\mathrm{Hz}$.

We find that the GR template minimizing the mismatch corresponds to a mass $M_{\rm best}\simeq 0.99\,M$, i.e.\ a $\sim 1\%$ shift relative to the true mass, this would predict a smaller memory as this scales with the total mass. 
Therefore, if we compare the amplitude of the memory between sGB and the GR waveform corresponding to $M_{\rm best}$ the final memory amplitude is of $\sim 4\%$ higher than in GR.
Ref.~\cite{Goncharov:2023woe} showed that Einstein Telescope could constrain the memory amplitude at the $\sim 2\%$ level by stacking all BBH events observed in one year. This suggests that the deviation found here may be within reach of next-generation detectors. However, that analysis does not account for the fact that, in sGB gravity, the relative deviation depends on the total mass of the system. A dedicated population analysis, similar to that performed in Ref.~\cite{Maselli:2023khq} for the ringdown, would therefore be required and we leave this for future work.

Using the mismatch results of Fig.~\ref{fig:mismatch_simple}, we can also estimate the number of events for which sGB and GR waveforms are distinguishable, following Ref.~\cite{Capuano:2026lhs}. Applying the indistinguishability criterion in Eq.~\eqref{eq:distcriter}, we find that for $M=20\,M_\odot$ the two waveforms are distinguishable up to redshift $z\sim 0.2$. To remain conservative, we adopt $D=9$ as the number of waveform parameters, as in Ref.~\cite{Julie:2024fwy}, although in our setup only the total mass is varied. While additional parameters could reduce the mismatch, our short waveforms greatly underestimate the total SNR and therefore the distinguishability horizon.

Assuming a merger rate density 
\[\sim \mathcal{O}(1\text{-}10)\,{\rm Gpc}^{-3}\,{\rm yr}^{-1}\,M_\odot^{-1}\] for BBHs with total mass $\sim 20\,M_\odot$~\cite{KAGRA:2021duu,LIGOScientific:2025pvj}, and a $\Lambda$CDM volume $V(z\leq0.2)\simeq 3\,{\rm Gpc}^3$, we obtain an expected rate of
\begin{equation}
N(z\leq0.2)\sim (0.3-3)\,{\rm yr}^{-1},
\end{equation}
Although only a fraction of these events will have nearly equal masses ($q\gtrsim 1.2$).
Overall, these results suggest that a few events with detectable deviations may be accessible to next-generation detectors, and that the inclusion of the memory could play a key role in breaking parameter degeneracies, as indicated by the increased mismatch when memory is included.

\section{Discussion and Conclusion}\label{sec:conclusion}

In this work, we have presented the first calculation of gravitational memory from full inspiral--merger--ringdown waveforms in a beyond-GR theory.
Based on the general formulas for displacement memory in Horndeski gravity \cite{Heisenberg:2023prj,Heisenberg:2024cjk,Maibach:2026wpz}, we derived the corresponding concrete expressions within a spin-weighted spherical harmonic expansion specialized to the case of sGB gravity. Our analysis captures, for the first time, the contribution from the merger phase --the most relevant for observational prospects-- where the bulk of the energy is emitted in both tensor and scalar radiation.

Our main phenomenological result is that the tensor memory in sGB gravity can differ measurably from its GR counterpart at the level of the dominant $(2,0)$ mode. For the configurations exhibiting the largest effect in our dataset, namely the shift-symmetric case with mass ratio $q\simeq 1.2$ and coupling $\lambda/m_2^2=0.1414$, we find that the memory amplitude of the sGB signal differs from the corresponding GR case by a few percent, and by up to $\sim 4\%$ when compared to the GR template with the total mass that minimizes the mismatch. Such a deviation could indeed be within reach of next-generation ground-based detectors~\cite{Goncharov:2023woe}. This deviation is driven primarily by the sGB-induced modification of the oscillatory tensor waveform during the late inspiral and merger, whereas the direct scalar contribution to the tensor memory remains suppressed by two to three orders of magnitude for the non-precessing binary black hole systems considered here. The dominant imprint of sGB gravity on the memory signal therefore arises indirectly, through the modified nonlinear merger dynamics encoded in the tensor radiation.

A second important result is that the inclusion of memory increases the minimum mismatch between GR and sGB waveforms, thereby improving their distinguishability in a detector-oriented comparison. In particular, for the representative high-coupling shift-symmetric configuration considered in our observability study, adding the memory raises the minimum mismatch by more than an order of magnitude relative to the oscillatory waveform alone. 
This indicates that memory can provide complementary information that helps break degeneracies between beyond-GR effects and source parameters, even if it is unlikely to constitute a standalone test of gravity. Owing to the limitations of our waveform model, these results rely primarily on variations of the total mass and the limited length of the signal. An important next step would be to investigate the impact of other source parameters, in particular the mass ratio and the spins, as well as the effect of a longer inspiral phase.

We note in particular that memory effects are especially promising for low-mass binaries observed by third-generation ground-based detectors such as the Einstein Telescope and Cosmic Explorer, since in this regime the memory signal falls within the most sensitive portion of the detector band. Consequently, gravitational-wave memory may provide a more powerful probe of deviations from GR than ringdown measurements, whose characteristic frequencies lie in the high-frequency regime where detector sensitivity is typically reduced. This motivates further investigation of memory in binary neutron star (NS) mergers~\cite{Barausse:2012da,Shibata:2013pra,Lam:2024azd,East:2022rqi,AresteSalo:2026zzc} and NS--BH systems in scalar-tensor theories~\cite{Corman:2024vlk}, where dipole radiation effects may be degenerate with the NS tidal deformability~\cite{Ma:2023sok}.


Our results are subject to several limitations. First, we restrict the computation of the memory to the dominant radiative modes, namely the $(2,2)$ mode in the tensor sector and the $(2,2)$ and $(1,1)$ modes in the scalar sector. Second, the numerical simulations provide the Weyl scalar $\Psi_4=\ddot h$, so that the strain must be reconstructed through two time integrations together with the usual extrapolation procedure. As recently emphasized \cite{Zosso:2026czc}, this standard reconstruction effectively removes the low-frequency content of the radiation \cite{Favata:2008yd,Favata:2010zu} and therefore yields a waveform corresponding to a \emph{memory-free} hypothesis, hence, a waveform without genuine low-frequency memory buildup. We therefore consistently add such a memory rise contribution in a post process by computing the Isaacson memory from the radiated energy flux, with an explicit spacetime averaging over the oscillatory scales of the primary radiation. In this way, obtain a controlled \emph{memory-full} signal model.

In GR, recent developments have made it possible to extract the full asymptotic waveform, including memory, using Cauchy characteristic extraction (CCE)~\cite{Bishop:1996gt,Mitman:2020pbt,Mitman:2024uss}, which solves the Einstein equations along null hypersurfaces up to future null infinity. To the best of our knowledge, such techniques have not yet been extended to beyond-GR theories. It would be interesting to test the current extrapolation-based method supplemented by the Isaacson memory computation against such a direct implementation of CCE, in particular in the context of the BMS balance laws as a diagnostic tool for waveform accuracy \cite{DAmbrosio:2024zok}. It is, however, expected that extrapolation-based waveforms capture high-frequency details quite accurately, while the low-frequency deficit is resolved via the Isaacson memory addition. Within an asymptotically flat spacetime context, a potential change in the propagation equation is furthermore not expected to influence the asymptotic region. This would, however, change significantly in an explicit cosmological context.



Exploring additional modified-gravity theories would further clarify the range of possible deviations in the memory beyond the amplitude enhancement found here. In this context, our results provide a starting point for theory informed parametrized tests of gravitational memory as an additional probe of deviations from GR, complementary to phase and ringdown tests. Overall, our findings suggest that gravitational memory may become a useful additional observable in future searches for strong-field deviations from GR, although a more complete assessment will require longer waveforms, broader parameter-space coverage, and a dedicated population analysis.



\section*{Acknowledgments}

This project has started and was developed within the memory subsection of the LISA Fundamental Physics Working Group.  
SG and LAS would like to thank the hospitality of the Centro de Ciencias de Benasque Pedro Pascual as well as the organizers and participants of the workshop ``New Frontiers in Strong Gravity'' in July 2024, where this work was conceived. We also acknowledge useful discussions with Marc Besançon, Henri Inchausp\'e, Lorena Magaña Zertuche and Thomas Sotiriou. We would like to thank David Trestini for his input to the draft. 
JZ is supported by funding from the Swiss National Science Foundation (Grant No. 222346) and the Janggen-Pöhn-Foundation. The Center of Gravity is a Center of Excellence funded by the Danish National Research Foundation under Grant No. 184. 
LAS is partly funded by
Interuniversitaire Bijzonder
Onderzoeksfonds (IBOF)/21/084.
DD acknowledges financial support from
the Spanish Ministry of Science and Innovation through the Ram\'on y Cajal programme (grant RYC2023-042559-I), funded by MCIN/AEI/10.13039/501100011033, from an Emmy Noether Research Group funded by the German Research Foundation (DFG) under Grant No. DO 1771/1-1,
and by the Spanish Agencia Estatal de Investigacion (grant PID2024-159689NB-C21) funded by MICIU/AEI/10.13039/501100011033 and by FEDER / EU. SY is supported by the European
Union-NextGenerationEU, through the National Recovery and Resilience Plan of the Republic of Bulgaria,
project No. BG-RRP-2.004-0008-C01. We acknowledge Discoverer PetaSC and EuroHPC JU for awarding this project access to Discoverer supercomputer resources.

\appendix
\onecolumngrid

\section{Formulas and Derivations}\label{App:Derivations}

\subsection{Spin-Weighted Spherical Harmonic Decomposition of Memory}\label{App:SWSHMemory}
In this appendix, we present the explicit derivation of the concrete SWSH expressions of the memory formula starting from Eq.~\eqref{NonLinDispMemoryModesSVT} with $\rho=1$ \cite{Heisenberg:2023prj}
\begin{equation}\label{eq:NonLinDispMemorysGBModesApp}
  \delta h_{lm}=r\sqrt{\frac{(l-2)!}{(l+2)!}}\int_{S^2}d^2\Omega'\,Y^*_{lm}(\Omega')\int_{-\infty}^u d u'\Big\langle\,|\dot{h}|^2+\dot\varphi_1^2\Big\rangle\,.
\end{equation}
Recall that the left hand side defines here the SWSH modes of the spin-weight $s=-2$ counterpart to Eq.~\eqref{eq:spinweitght2scalar} for the memory contribution [Eq.~\eqref{eq:SW Memory quantity}]
\begin{equation}\label{eq:SW Memory quantity GR App}
   \delta h(u,r,\Omega)\equiv \delta h_{ij}^\text{TT}\bar m^i\bar m^j= \delta h_+-i \delta h_\times=\sum_{l=2}^\infty\sum_{m=-l}^{l}\, \delta h_{lm}(u,r)\,\,_{\mys{-2}}Y_{lm}(\Omega)\,,
\end{equation}
where the complex transverse vector of spin-weight $s=-1$ is defined as
\begin{equation}
    \bar m_i\equiv ({1}/{\sqrt{2}})(u_i-iv_i)\,.
\end{equation}
Note that $_{\mys{-2}}Y_{lm}(\Omega)=0$ for $l<2$, such that by construction such low $l$ modes are unphysical. 

\subsubsection{Analytic angular integration}
Within the memory formula in Eq.~\eqref{eq:NonLinDispMemorysGBModesApp}, the angular integral can be performed analytically by also explicitly expanding the primary waves into their SWSH modes as defined in Eqs.~\eqref{eq:SW Memory quantity GR} and \eqref{eq:SW Memory quantity scalar}. Indeed, we have that
\begin{equation}
    |\dot h|^2=\sum_{l_1=2}^\infty\sum_{m_1=-l_1}^{l_1}\sum_{l_2=2}^\infty\sum_{m_2=-l_2}^{l_2} \dot h_{l_1m_1}\dot h^*_{l_2m_2}\,_{\mys{-2}}Y_{l_1m_1}\,_{\mys{-2}}Y^*_{l_2m_2}\,,
\end{equation}
while
\begin{equation}
    \dot \varphi^2=\sum_{l_3=0}^\infty\sum_{m_3=-l_3}^{l_3}\sum_{l_4=0}^\infty\sum_{m_4=-l_4}^{l_4} \dot \varphi_{l_3m_3}\dot \varphi^*_{l_4m_4}\,Y_{l_3m_3}\,Y^*_{l_4m_4}\,,
\end{equation}
such that using the rule for complex conjugation of the SWSHs
\begin{equation}
    (-1)^{s+m}\, \phantom{}_{\mys{-s}}Y^*_{l-m}=\phantom{}_{\mys s}Y_{lm}\,,\label{CCSWSH H}
\end{equation}
one obtains
\begin{align}\label{eq:NonLinDispMemoryModessGB 2}
\delta h_{lm}=&(-1)^{m}r\sqrt{\frac{(l-2)!}{(l+2)!}}
\int_{-\infty}^u d u'\Bigg[\sum_{l_3m_3l_4m_4}(-1)^{m_4}\big\langle \dot \varphi_{l_3m_3}\dot \varphi^*_{l_4m_4}\big\rangle \int_{S^2} d^2\Omega'\,\,Y_{l_3m_3}\,Y_{l_4-m_4} \,Y_{l-m}\nonumber \\
  &+\sum_{l_1m_1l_2m_2}(-1)^{m_2}\big\langle \dot h_{l_1m_1}\dot h^*_{l_2m_2}\big\rangle \int_{S^2} d^2\Omega'\,\,_{\mys{-2}}Y_{l_1m_1}\,_{\mys{2}}Y_{l_2-m_2} \,Y_{l-m}\Bigg]\,.
\end{align}
The angular integrals appearing in the equation above indeed have a known analytic solution, since a triple integral of SWSHs can be written as
\begin{equation}\label{SWSHTrippleInt}
    \int_{S^2}d^2\Omega\,_{\mys{s'}}Y_{l'm'}\,_{\mys{s''} }Y_{l''-m''}\,_{\mys{s}}Y_{l-m}
    = S^{l'l''l}
\begin{pmatrix}
l' & l'' & l\\
m' & -m'' & -m
\end{pmatrix}
\begin{pmatrix}
l' & l'' & l\\
-s' & -s'' & -s
\end{pmatrix}\,,
\end{equation}
with
\begin{equation}
   S^{l'l''l}\equiv \sqrt{\frac{(2l'+1)(2l''+1)(2l+1)}{4\pi}}\,,
\end{equation}
provided that $s'+s''+s=0$\footnote{If the condition $s'+s''+s=0$ is not satisfied, the relation in Eq.~\eqref{SWSHTrippleInt} is not valid and the result can be expressed in terms of gamma functions instead, as for instance presented in the appendix of \cite{Favata:2008yd}.}. Here, the round brackets define Wigner $3-j$ symbols and the integral is therefore only non-zero if 
\begin{equation}\label{eq:SelectionRule}
    m=m'-m''+m \quad \text{and} \quad |l'-l''|\leq l\leq l'+l''\,.
\end{equation}
Given a spin-weighted mode-decomposition of the leading order high-frequency wave, the full expression of the memory component in sBG can therefore be written as
\begin{align}\label{eq:NonLinDispMemoryModessBG 3}
  \delta h_{lm}=&\,r\Bigg[\sum_{l_1l_2\geq 2}\sum_{m_1m_2}\Gamma^{l_1m_1l_2m_2lm}_{-2,2,0} \int_{-\infty}^ud u'\big\langle \dot h_{l_1m_1}\dot h^*_{l_2m_2}\big\rangle+\sum_{l_3l_4\geq 0}\sum_{m_3m_4}\Gamma^{l_3m_3l_4m_4lm}_{0,0,0} \int_{-\infty}^ud u'\big\langle \dot \varphi_{l_3m_3}\dot \varphi^*_{l_4m_4}\big\rangle\Bigg]\,,
\end{align}
with
\begin{equation}\label{eq:Gamma}
    \Gamma^{l'm'm''l''lm}_{s',s'',s}\equiv (-1)^{m+m''}\sqrt{\frac{(l-2)!}{(l+2)!}} \,S^{l'l''l}\,
    \begin{pmatrix}
        l' & l'' & l\\
         m' & -m'' & -m
    \end{pmatrix}
    \begin{pmatrix}
        l' & l'' & l\\
        -s' & -s'' & -s
    \end{pmatrix}.
\end{equation}
Note that this formula is only valid for $l\geq 2$ and $\Gamma^{l'm'm''l''lm}_{s',s'',s}$ should be set to zero otherwise.

\subsubsection{Leading Order Memory for a Non-Precessing CBC} 
Assuming the particular case of non-precessing CBC events, Eq.~\eqref{eq:NonLinDispMemoryModessBG 3} can efficiently be evaluated. 
For a non-precessing binary systems in a frame where the orbital plane coincides with the $x-y$ plane, the high-frequency waves enjoy a symmetry under reflection across the orbital plane 
\begin{equation}
    h(u,r,\theta,\phi)=h^*(u,r,\pi-\theta,\phi)\,,\quad \varphi(u,r,\theta,\phi)=\varphi^*(u,r,\pi-\theta,\phi)\,,
\end{equation}
that leads to the following relation between the corresponding modes \cite{Zosso:2024xgy}
    \begin{equation}\label{eq:SymmOrbitPlaneModes}
         h_{lm}=(-1)^l\,h^*_{l-m}\,,\quad  \varphi_{lm}=(-1)^l\,\varphi^*_{l-m}\,.
    \end{equation}
Moreover, for a non-precessing binary merger, gravitational memory is dominated by its $(2,0)$ and $(4,0)$ modes, as shown for instance in Refs.~\cite{Favata:2009ii,Zosso:2026czc}. This is imposed by the averaging $\langle ... \rangle$ over the oscillatory time-scales and the fact that the asymptotic field modes follow the orbital phase $\phi(u')$ as $\propto e^{-im'\phi(u')}$. Plugging this overall behavior into  Eq.~\eqref{eq:NonLinDispMemoryModessBG 3} we obtain due to Eq.~\eqref{eq:SelectionRule} that
\begin{equation}\label{eq:phaseBehaviorM}
     \delta h_{l m}\propto \big\langle e^{-i(m'-m'')\phi(u')}\big\rangle= \big\langle e^{-im\phi(u')}\big\rangle\,.
\end{equation}
and all modes with $m\neq 0$ will be heavily suppressed due to the averaging.
    
Finally, for non-precessing CBCs the $(2,2)$ mode of the primary tensor waves dominates, while for unequal mass ratios the scalar radiation possesses a dominant $(1,1)$ dipole radiation with a subdominant $(2,2)$ mode \cite{Yagi:2011xp,Corman:2022xqg}. Putting everything together, the dominant memory modes will therefore be given by 
\begin{equation}
      \delta h_{20}=\frac{r}{7}\sqrt{\frac{5}{6\pi}}  \int_{-\infty}^u d u'\big\langle |\dot h_{22}|^2-|\dot \varphi_{22}|^2\big\rangle-\frac{r}{2\sqrt{30\pi}}  \int_{-\infty}^u d u'\big\langle |\dot \varphi_{11}|^2\big\rangle\,,
\end{equation}
and
\begin{equation}
    \delta h_{40} =\frac{r}{1260}\sqrt{\frac{5}{2\pi}}  \int_{-\infty}^u d u'\big\langle |\dot h_{22}|^2+6\,|\dot \varphi_{22}|^2\big\rangle\, ,
\end{equation}
due to the selection rule in Eq.~\eqref{eq:SelectionRule}, while Eq.~\eqref{eq:SymmOrbitPlaneModes} ensures that $\delta h_{30}=0$. 
Due to the change in the asymptotic scalar charge as the binary merges into a single remnant, the scalar monopole $\varphi_{00}$ also acquires a considerable amplitude.
However, again due to the selection rule in Eq.~\eqref{eq:SelectionRule} the
monopole scalar modes do not contribute to the tensor memory.

Plugging the results into the expansion in Eq.~\eqref{eq:SW Memory quantity GR} finally yields
\begin{align}
    \delta h_+(u,r,\theta,\phi)=&~\frac{r}{192\pi}\,\sin^2\theta(17+\cos^2\theta)\,\int_{-\infty}^u d u'\big\langle |\dot h_{22}|^2\big\rangle-\frac{r}{32\pi}\,\sin^2\theta(3-\cos^2\theta)\,\int_{-\infty}^u d u'\big\langle|\dot \varphi_{22}|^2\big\rangle\nonumber\\
    &-\frac{r}{16\pi}\sin^2\theta\int_{-\infty}^u d u'\big\langle |\dot \varphi_{11}|^2\big\rangle\,, \label{MemQuadrupole}\\
     \delta h_\times(u,r,\theta,\phi)=&~0\,.
\end{align}
Thus, notably, in the given coordinate system, the memory signal only has a single polarization. 
Moreover, observe that Eq.~\eqref{MemQuadrupole} also implies that edge-on systems with $\theta=\pi/2$ are most optimal for memory detection, while the memory effect in the above approximation vanishes in the face-on limit $\theta=0$.

For an equal mass ratio, however, the dipole radiation of the scalar field vanishes, since in this case the spacetime is even under parity inversion $n_i\rightarrow -n_i$, while the spherical harmonics of any odd multipole of the scalar are odd \cite{Corman:2022xqg}. Thus, in this case the dominant scalar mode is the $(2,2)$ mode and the leading order contribution to the memory reads
\begin{equation}
     \delta h_{20} =\frac{r}{7}\sqrt{\frac{5}{6\pi}}  \int_{-\infty}^u d u'\big\langle |\dot h_{22}|^2-|\dot \varphi_{22}|^2\big\rangle\,,
\end{equation}
and
\begin{equation}
    \delta h_{40} =\frac{r}{1260}\sqrt{\frac{5}{2\pi}}  \int_{-\infty}^u d u'\big\langle |\dot h_{22}|^2+6\,|\dot \varphi_{22}|^2\big\rangle\,.
\end{equation}

\section{The total displacement memory offset}\label{App:BalanceLaws}

In this appendix, we summarize how the total displacement-memory offset in sGB gravity follows from the generalized BMS flux-balance laws, and we clarify its relation to the Isaacson null-memory formulas used in the main text. In particular, this allows us to separate the total memory into its null and ordinary contributions and to justify why, for the non-precessing compact binary coalescences considered here, the ordinary-memory terms are negligible.

\subsection{BMS balance laws in sGB gravity}

In a post-Newtonian Bondi frame, in which 
\begin{equation}
    \lim_{u_0\rightarrow-\infty}h(u_0)=0\,,
\end{equation}
the BMS flux-balance laws can generically be written as
\begin{align}\label{equ:BMSSupermomentumBalanceLawBD2}
  \oint_{S^2} d^2\Omega\,\alpha(\theta,\phi)\,\text{Re}[\eth^2 h(u)]&= \mathcal{P}_{\alpha(\theta,\phi)}(u)- 4 \Big(\mathcal Q_{\alpha(\theta,\phi)}(u)-\mathcal Q^0_{\alpha(\theta,\phi)}\Big)\,,
\end{align}
Here, the BMS balance laws express the non-conservation of asymptotic charges at null infinity by relating the change in a given supermomentum charge $\mathcal Q_{\alpha(\theta,\phi)}(u)$ to the flux of radiative energy-momentum through null infinity $\mathcal{P}_{\alpha(\theta,\phi)}(u)$, up to the supertranslation-dependent contribution on the left-hand side. The function $\alpha=\alpha(\theta,\phi)$ labels the corresponding BMS supertranslation on the celestial two-sphere, which can be thought of as an extension of Poincar\'e translations to angle-dependent translations. As we will explicitly see below, the left-hand side vanishes for the lowest spherical harmonics, and the associated charges reduce to the ordinary Bondi four-momentum. On the other hand, generic choices of $\alpha$ define the higher supermomentum charges, for which the left-hand side is non-trivial and, in the infinite-time limit, gives rise to gravitational memory.

In sGB gravity, the null flux associated with the supertranslation generated by $\alpha(\theta,\phi)$ reads \cite{Maibach:2026wpz}
\begin{align}\label{equ:PPPsGB}
    \mathcal{P}^\text{sGB}_{\alpha(\theta,\phi)}(u)\equiv \int_{-\infty}^u\dd u'\int\dd^2\boldsymbol\Omega\, \,\alpha(\theta,\phi) \Big[|\dot{h}|^2+ \dot{\varphi}_1^{\,2}\Big]\,.
\end{align}
On the other hand, the supermomentum charge at each retarded time $u$ is given by
\begin{equation}
    \mathcal Q^\text{sGB}_{\alpha(\theta,\phi)}(u)=\oint_{S^2} \dd^2\boldsymbol\Omega\,\alpha(\theta,\phi)\, \mathscr M_{\rm sGB}(u,\theta,\phi)
\end{equation}
with
\begin{equation}
     \mathscr M_{\rm sGB}= \mathscr M_{\rm GR}-\frac{1}{4}\varphi_1\dot{\varphi}_1\,,
\end{equation}
where $ \mathscr M_{\rm GR}(u,\theta,\phi)$ is the GR Bondi mass aspect, and where we also define the initial supermomentum charge
\[
\mathcal Q^0_{\alpha(\theta,\phi)}\equiv   \lim_{u_0\rightarrow-\infty}\mathcal Q^\text{sGB}_{\alpha(\theta,\phi)}(u_0)\,.
\]

In these balance laws, the total strain scalar $h$ [Eq.~\eqref{eq:spinweitght2scalar}] on the left-hand side of Eq.~\eqref{equ:BMSSupermomentumBalanceLawBD2} can be isolated by expanding it in terms of SWSHs as in Eq.~\eqref{eq:SW Memory quantity GR}. Using the identity
\begin{align}
    \eth^2\,{}_{\scriptscriptstyle -2}Y_{lm}
    &=\sqrt{\frac{(l+2)!}{(l-2)!}}\,Y_{lm}\,,
\end{align}
as well as choosing $\alpha(\theta,\phi)=Y^*_{lm}(\theta,\phi)$, the orthogonality relations of the spherical harmonics imply that the left-hand side of Eq.~\eqref{equ:BMSSupermomentumBalanceLawBD2} naturally projects onto the electric-parity component of the spin-weighted strain
\begin{equation}\label{eq:AHlmToUlmVlm}
    h_{lm}=\frac{1}{\sqrt{2}}\left[h^E_{lm}-ih^M_{lm}\right]\,,
\end{equation}
which can be inverted to yield
\begin{align}
    h^E_{lm}
    =\frac{1}{\sqrt{2}}
    \left[h_{lm}+(-1)^m h^*_{l,-m}\right]\,.
    \label{eq:AUVlmToHlm}
\end{align}
Hence, one ends up with
\begin{align}\label{BMSSupermomentumBalanceLawMemoryE}
   h^E_{lm}(u)
  =\sqrt{\frac{2(l-2)!}{(l+2)!}}
  \Big[\mathcal{P}_{Y^*_{lm}}(u)-4\,\Big(\mathcal Q_{Y^*_{lm}}(u)-\mathcal Q^0_{Y^*_{lm}}\Big)\Big]\,.
\end{align}
Note that this equation represents a constraint equation for the full asymptotic tensor strain $h$, including all oscillatory gravitational waves.

Note that Eq.~\eqref{BMSSupermomentumBalanceLawMemoryE} is only valid for $l\geq 2$, since
\begin{equation}
    \sqrt{\frac{(l+2)!}{(l-2)!}}=\sqrt{(l+2)(l+1)l(l-1)}\,.
\end{equation}
Instead, for $l=0$ and $l=1$, the left-hand side of Eq.~\eqref{equ:BMSSupermomentumBalanceLawBD2} vanishes and the balance laws describe the Bondi mass-loss and momentum-loss formulas. More precisely, for $l=0$ one has 
\begin{equation}
 \mscr_{\text{GR}}(u)- \mscr^0_{\text{GR}} = \frac{1}{4} \int_{-\infty}^{u} du' \oint d^2\Omega\left( |\dot{h}|^2 + \dot{\varphi}_1^{\,2} \right)  + q(u)-q^0
\end{equation}
where we define the GR Bondi mass 
\begin{equation}
 \mscr_{\text{GR}}(u)\equiv \oint_{S^2} d^2\Omega\, \mathscr M_{\rm GR}(u,\theta,\phi)\,,
\end{equation}
and equivalently
\begin{equation}
    q(u)\equiv \frac{1}{4}  \oint  d^2\Omega \,\varphi_1 \dot{\varphi}_1\,.
\end{equation}

\subsection{Total null and ordinary memory}

The total gravitational memory offset
\begin{equation}
    \Delta h\equiv \lim_{u\to\infty}h(u)-h(-\infty)
\end{equation}
is extracted by taking the limit $u\to \infty$, hence
\begin{align}\label{BMSSupermomentumBalanceLawMemory}
   \Delta h_{lm}
  =\sqrt{\frac{(l-2)!}{(l+2)!}}
  \Bigg[\int_{-\infty}^\infty\dd u'\int\dd^2\boldsymbol\Omega\, \,Y^*_{lm} \Big(|\dot{h}|^2+ \dot{\varphi}_1^{\,2}\Big)-4\,\Delta\mathcal Q_{Y^*_{lm}}\Bigg]\,.
\end{align}
To write this equation, we have used the fact that there is no memory offset in the magnetic-parity component of the asymptotic strain $h^M_{lm}$ and Eq.~\eqref{eq:AHlmToUlmVlm}.

The right-hand side can be further simplified by expanding all fields entering $\mathcal{P}$ and $\mathcal Q$ in SWSHs, allowing the angular integrals to be evaluated analytically. Concretely,
\begin{align}\label{BMSSupermomentumBalanceLawBD3}
   \mathcal Q^\text{sGB}_{Y^*_{lm}}=&\oint_{S^2} \dd^2\boldsymbol\Omega\,Y^*_{lm}\sum_{l', m'}\sum_{l'', m''}\left(  \mathscr M_{{\rm GR},\,l'm'}Y_{l'm'}-\frac{1}{4}\varphi_{l'm'}\dot\varphi_{l''m''} Y_{l'm'}Y_{l''m''}\right)\nonumber\\
  &= \mathscr M_{{\rm GR},\,lm} - \frac{1}{4}\sum_{l', m'} \sum_{l'', m''} \Gamma_{0, 0, 0}^{l' m' m'' l'' l m}  \varphi_{l', m'} \dot{\varphi}^*_{l'', m''}\,,
\end{align}
where the $\Gamma$ factor is defined in Eq.~\eqref{eq:Gamma}. Together with the result in Eq.~\eqref{eq:NonLinDispMemoryModessGB 2}, the full BMS balance laws of sGB gravity can therefore be written as
\begin{align}\label{eq:TotalMemoryOffsetBalanceSWSH}
\Delta h_{lm} = \sum_{l', m'} \sum_{l'', m''} \int_{-\infty}^{\infty} du' \Big[ &\Gamma_{-2, 2, 0}^{l' m' m'' l'' l m} \dot{h}_{l', m'} \dot{h}^*_{l'', m''} + \Gamma_{0, 0, 0}^{l' m' m'' l'' l m} \dot{\varphi}_{l', m'} \dot{\varphi}^*_{l'', m''} \Big] \nonumber \\ 
&- 4 \Delta \mathscr M_{{\rm GR},\,lm} + \sum_{l', m'} \sum_{l'', m''} \Gamma_{0, 0, 0}^{l' m' m'' l'' l m} \Delta (\varphi_{l', m'} \dot{\varphi}^*_{l'', m''})\,.
\end{align}

It is interesting to compare these total memory-offset equations in Eqs.~\eqref{BMSSupermomentumBalanceLawMemory} and \eqref{eq:TotalMemoryOffsetBalanceSWSH}, obtained from the balance laws, with the null-memory formulas in Eqs.~\eqref{eq:NonLinDispMemorysGBModesApp} and \eqref{eq:NonLinDispMemoryModessBG 3}, obtained via the Isaacson approach. First of all, in contrast to the total memory offset accessed through the BMS balance laws, the Isaacson formulas in Eqs.~\eqref{eq:NonLinDispMemorysGBModesApp} and \eqref{eq:NonLinDispMemoryModessBG 3} describe the time-dependent memory rise signal. This is achieved via an explicit averaging over oscillatory variation scales, which allows for the extraction of a conservative definition of a memory signal at each instant of time of a given event \cite{Zosso:2026czc}.

Moreover, while the flux balance laws define the total memory offset, the Isaacson formulas in Eqs.~\eqref{eq:NonLinDispMemorysGBModesApp} and \eqref{eq:NonLinDispMemoryModessBG 3} focus on the null-memory contribution. In GR, the additional ordinary-memory component is associated with so-called linear memory, which arises due to unbound matter sources as well as black-hole remnant kicks. These can also be computed within the Isaacson framework based on the bulk matter energy-momentum tensor, as discussed in Ref.~\cite{Heisenberg:2024cjk}. However, for non-precessing CBC events, the ordinary-memory contributions are generally subdominant. Indeed, in the absence of unbound matter, remnant-kick contributions from non-precessing binaries remain negligible \cite{Christodoulou:1991cr,Favata:2008ti,Tahura:2021hbk}, while the additional scalar term in Eq.~\eqref{eq:TotalMemoryOffsetBalanceSWSH} vanishes for localized events, since $\dot\varphi_1 \approx 0$ for stationary initial and final configurations.

\section{Mismatch for dynamical scalarization case}\label{app:mismatchDS}

\begin{figure}[H]
    \centering
    \includegraphics[width=0.5\linewidth]{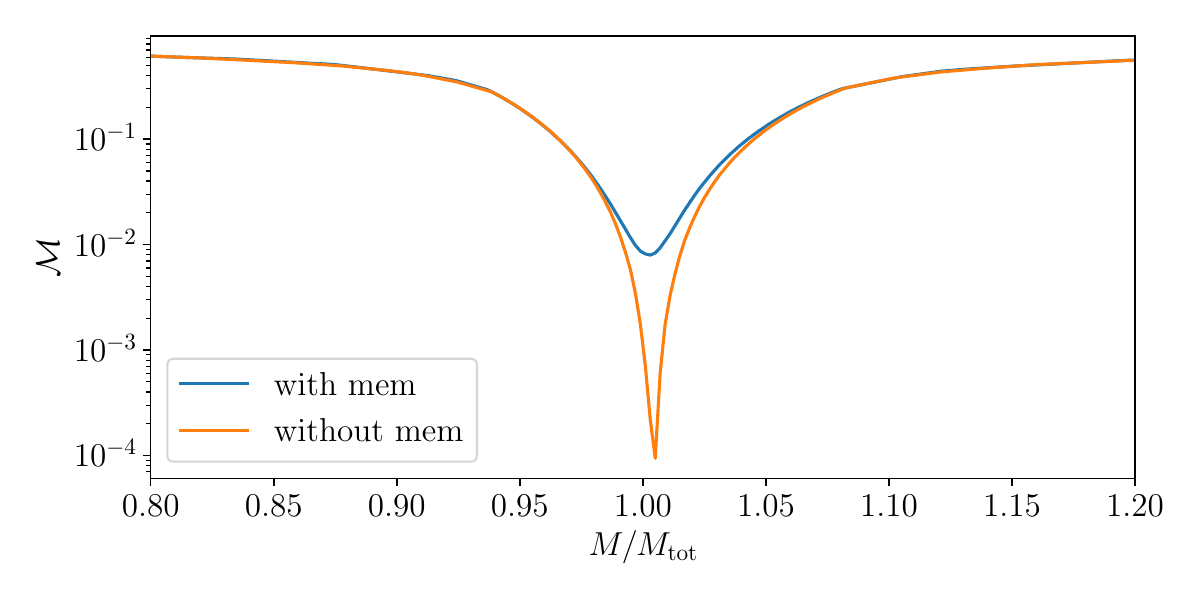}
    \caption{Mismatch results similar to Fig.~\ref{fig:mismatch_simple} for the waveforms in Fig.~\ref{fig:DSwaveforms} corresponding to the dynamical scalarization case with equal masses. }
    \label{fig:mismatchDS}
\end{figure}

In Fig.~\ref{fig:mismatchDS}, we show the mismatch for the waveforms described in Sec.~\ref{sec:synscal}, computed for $\beta=4$, total mass $M=20M_\odot$, and inclination $\iota=90^\circ$. As in Sec.~\ref{sec:observability}, we compare sGB and GR waveforms using the dominant oscillatory $(2,2)$ mode, both with and without the inclusion of memory. 
We find a mismatch of $\mathcal{M}=9\times 10^{-5}$ when memory is neglected, which increases to $\mathcal{M}=8\times 10^{-3}$ when memory is included. Consistently with the shift-symmetric case, the inclusion of memory enhances the mismatch, thereby improving the distinguishability between the two models. 
However, in this case we do not observe a significant shift in the mass that minimizes the mismatch, as the relative shift remains small, $M/M_{\rm tot}-1 \lesssim 10^{-2}$. This is expected because of the smaller difference between sGB and GR in this case, which also reflects in the smaller modification of the tensor memory.

\twocolumngrid
\newpage
\bibliographystyle{utcaps}
\bibliography{refs}

\clearpage

\end{document}